\documentclass[AMA,STIX1COL]{WileyNJD-v2}

\usepackage{moreverb}
\usepackage{color, soul} %for highlighting
\usepackage{graphicx}
\usepackage{url}  
\usepackage{amsthm,amsmath,bm,mathtools}
\usepackage{dsfont}
\usepackage{multirow}
\usepackage{enumerate}
\usepackage{booktabs}
\usepackage{comment}
\usepackage{algorithm,algpseudocode} %for algorithm
\usepackage[font=normalsize,skip=0pt]{caption}
\newcommand{\code}[1]{\texttt{#1}}
\newcommand{\R}{{\normalfont\textsf{R }}{}}
\newcommand{\lsq}{\left[}
\newcommand{\rsq}{\right]}
\newcommand{\lbc}{\left \{ }
\newcommand{\rbc}{\right \} }
\newcommand{\lp}{\left(}
\newcommand{\rp}{\right)}

\newcommand{\cond}{{\, \vert \,}}

\makeatletter
\g@addto@macro\normalsize{%
  \setlength\abovedisplayskip{5pt}
  \setlength\belowdisplayskip{5pt}
  \setlength\abovedisplayshortskip{1pt}
  \setlength\belowdisplayshortskip{1pt}
} \makeatother
\makeatletter
\newcommand{\distas}[1]{\mathbin{\overset{#1}{\kern\z@\sim}}}%
\newsavebox{\mybox}\newsavebox{\mysim}
\newcommand{\distras}[1]{%
  \savebox{\mybox}{\hbox{\kern3pt$\scriptstyle#1$\kern3pt}}%
  \savebox{\mysim}{\hbox{$\sim$}}%
  \mathbin{\overset{#1}{\kern\z@\resizebox{\wd\mybox}{\ht\mysim}{$\sim$}}}%
}
\makeatother

\articletype{Article Type}%

\received{<day> <Month>, <year>}
\revised{<day> <Month>, <year>}
\accepted{<day> <Month>, <year>}

\begin{document}

\title{A flexible approach for causal inference with multiple treatments and clustered survival outcomes}

\author[1]{Liangyuan Hu*}

\author[1]{Jiayi Ji}

\author[2]{Ronald D. Ennis}

\author[3]{Joseph W. Hogan}

\authormark{Hu \textsc{et al}}

\address[1]{\orgdiv{Department of Biostatistics and Epidemiology}, \orgname{Rutgers University}, \orgaddress{\state{Piscataway, NJ 08854}, \country{U.S.A.}}}

\address[2]{\orgdiv{Department of Radiation Oncology}, \orgname{Cancer Institute of New Jersey of Rutgers University}, \orgaddress{\state{New Brunswick, NJ 08903}, \country{U.S.A.}}}

\address[3]{\orgdiv{Department of Biostatistics}, \orgname{Brown University}, \orgaddress{\state{Providence, RI 02912}, \country{U.S.A.}}}

\corres{*Liangyuan Hu, PhD\\
683 Hoes Lane West, Piscataway, NJ 08854\\ \email{liangyuan.hu@rutgers.edu}}

\abstract[Abstract]{When drawing causal inferences about the effects of multiple treatments on clustered survival outcomes using observational data, we need to address implications of the multilevel data structure, multiple treatments, censoring and unmeasured confounding for causal analyses. 
Few off-the-shelf causal inference tools are available to simultaneously tackle these issues. We develop a flexible random-intercept accelerated failure time model, in which we use Bayesian additive regression trees  to capture arbitrarily complex relationships between censored survival times and pre-treatment covariates and use the random intercepts to capture cluster-specific main effects. We develop an efficient Markov chain Monte Carlo algorithm to draw posterior inferences about the population survival effects of multiple treatments and examine the variability in cluster-level effects. We further propose an interpretable sensitivity analysis approach to evaluate the sensitivity of drawn causal inferences about treatment effect to the potential magnitude of departure from the causal assumption of no unmeasured confounding. Expansive simulations empirically validate and demonstrate good practical operating characteristics of our proposed methods. Applying the proposed methods to a dataset on older high-risk localized prostate cancer patients drawn from the National Cancer Database, we evaluate the comparative effects of three treatment approaches on patient survival, and assess the ramifications of potential unmeasured confounding. The methods developed in this work are readily available in the $\R$ package $\textsf{riAFTBART}$.}


\jnlcitation{\cname{%
\author{Hu L}, 
\author{Ji J}, 
\author{Ennis D. R}, 
\author{Hogan W. J}} (\cyear{2022}), 
\ctitle{A flexible approach for causal inference with multiple treatments and clustered survival outcomes}, \cjournal{Statistics in Medicine}, \cvol{2022;00:1--18}.}

\maketitle

%\footnotetext{\textbf{Abbreviations:} ANA, anti-nuclear antibodies; APC, antigen-presenting cells; IRF, interferon regulatory factor}

\section{Introduction}\label{sec1}

 In cancer research,  decision makers are starting to rely more heavily on real world evidence because clinical trials can be enormously expensive, time consuming and restrictive. The increasing availability of observational data sources like large registries and electronic health records provides new opportunities to obtain real world comparative effectiveness evidence. Recent efforts have been made to evaluate the comparative effectiveness of multiple treatment approaches on patient survival for high-risk localized prostate cancer using the large-scale national cancer database (NCDB). \citep{ennis2018brachytherapy,chen2018challenges,hu2021estimating,zeng2022propensity} The complex data structures, however, pose three main challenges for statistical analyses that have not been fully addressed in the extant literature. 
 
 First, 
 %there are three types of treatments for high-risk localized prostate cancer: radical prostatectomy, external beam radiotherapy (EBRT) combined with androgen deprivation (AD) and EBRT plus brachytherapy. 
 multiple (i.e., more than two) active treatments are involved. For high-risk localized prostate cancer, there are three popular treatment approaches: radical prostatectomy (RP), external beam radiotherapy (EBRT) combined with androgen deprivation (AD) (EBRT+AD) and EBRT plus brachytherapy with or without AD (EBRT+brachy$\pm$AD). Each treatment option has historically been performed on different types of patients as various health and demographic factors are closely linked to treatment choice.  %Long-term survival is an important utility for comparing the effectiveness of the three treatments. 
 Second, treatment and patient information on a large, national population is collected from various treating facilities. The NCDB hospital registry data are collected in more than 1,500 Commission on Cancer accredited facilities. The participating institutions are not selected at random, and there can be substantial institutional variation in treatment effect.  Third, some important confounders (pre-treatment variables predicting both treatment and outcome) may not be collected in the observational data. Two known main
confounders in high-risk prostate cancer are the number of positive cores and magnetic resonance imaging findings, which are often not collected in the NCDB data or other large cancer registries alike. 
 
Despite numerous recent advances in causal inference, the literature for handling data
structures of this type – which arise frequently in population cancer research – is sparse. Causal inference techniques traditionally focus on a binary treatment.  There is now a substantial body of research on causal inference methods with multiple treatments and a continuous outcome \citep{feng2012generalized, mccaffrey2013tutorial, linden2016estimating} or a binary outcome. \citep{hu2020estimation,hu2021estimation,yu2021comparison}  Ennis et al. \citep{ennis2018brachytherapy} and Zeng et al. \citep{zeng2022propensity} described propensity score weighting based methods for drawing inferences about causal effects of multiple treatments on censored survival outcomes, but neither work considered the multilevel data structure. There are two main reasons why it might be important to account for the cluster-level, or institutional variation when estimating  treatment effect in the general population. First,  if there are substantial institutional effects, then a causal analysis ignoring institution would be based on an incorrect model, which can lead to invalid inferences about treatment effects. Second, neither participating institutions nor patients in the registry data were selected at random. With substantial institutional variation, it is unclear exactly what treatment effect would be seen in the general patient population across various institutions with different outcomes. %it would be more like that seen in the institutions with poor outcomes or that seen in those with good outcomes. 
Finally, inferring causal links from observational data inevitably involves the untestable assumption of no unmeasured confounding, which holds that all pre-treatment variables are sufficient to predict both treatment and outcome. If there are unmeasured confounders, it is important to evaluate  how departures from the no unmeasured confounding assumption might alter causal conclusions. Sensitivity analysis is useful to address this causal assumption, and is recommended by the Strengthening the Reporting of Observational Studies in Epidemiology (STROBE) guidelines. \citep{von2007strengthening} Existing sensitivity analysis approaches have largely focused on a binary treatment.  Hu et al. \citep{hu2022flexible} recently proposed a flexible Monte Carlo sensitivity analysis approach in the context of multiple treatments and a binary outcome. There is still sparse literature on sensitivity analysis methods that simultaneously accommodate multiple treatments and multilevel censored survival outcomes.  

To fill these research gaps, we propose a flexible approach for drawing causal inferences about multiple treatments while respecting the multilevel survival data structure, and develop an interpretable sensitivity analysis approach to evaluate how the drawn causal conclusions might  be altered in response to the potential magnitude of departure from the no unmeasured confounding assumption. We propose a random-intercept accelerated failure time (AFT)  model utilizing Bayesian additive regression trees (BART) \citep{chipman2010bart}, termed as riAFT-BART. In this model, we use the random intercepts for  cluster-specific main effects capturing the variation across the institutions, and leverage the flexibility of the BART model \citep{hu2020estimation,hill2011bayesian, hu2021variable,hu2021est} to accurately capture arbitrarily complex relationships among survival times, treatments and covariates. We then develop an efficient Markov chain Monte Carlo algorithm to draw posterior inferences about the population survival effects of multiple treatments. We further propose an interpretable sensitivity analysis approach leveraging our riAFT-BART model in the context of multiple treatments and clustered survival outcomes.  Finally, we apply the proposed methods to the NCDB  data and elucidate the causal effects of three treatment approaches (RP, EBRT+AD and EBRT+brachy$\pm$AD) on patient survival among older and high-risk prostate cancer patients and the impact of unmeasured confounding, as well as examine the institutional effects. %We will further perform a sensitivity analysis to evaluate how sensitive the estimated causal effects are to possible unmeasured confounding. 

The rest of the paper is organized as follows. Section~\ref{sec:est_methods} describes notation and proposes the riAFT-BART method for the estimation of causal effects. Section~\ref{sec:sa_methods} describes a corresponding sensitivity analysis approach using the model developed in Section~\ref{sec:est_methods}. Section~\ref{sec:sim} develops a wide variety of simulation scenarios to examine the practical operating characteristics of our proposed methods against three alternative methods, and presents findings.  In Section~\ref{sec:application}, we apply our methods to NCDB data to estimate the causal effects of three treatment approaches on patient survival among high-risk localized prostate cancer patients, and perform a sensitivity analysis to evaluate how sensitive the estimated causal effects are to possible unmeasured confounding.  Section~\ref{sec:disc} concludes with a discussion. 

\section{Estimation of causal effects} \label{sec:est_methods}
\subsection{Notation, definitions and assumptions}
Consider an observational study possessing a two-level data structure that has $K$ clusters (institutions), each having treated $n_k$ individuals, indexed by $i=1, \ldots, n_k, k=1, \ldots, K$.  The total number of individuals in the study is $N = \sum_{k=1}^K n_k$.  Our goal is to infer the causal effect of treatment $A \in \mathscr{A} = \{a_1, \ldots,a_J\}$  on time to failure $T$, where $J$ is the total number of treatment options. For each individual $i$ in cluster $k$, there is a vector of pre-treatment measured covariates $\bm{X}_{ik}$, and let $T_{ik}$ be the individual's  failure time, which may be right censored at $C_{ik}$. The observed outcome consists of $Y_{ik}=\min (T_{ik}, C_{ik})$ and the censoring indicator $\Delta_{ik}=I (T_{ik}<C_{ik})$. Let $V_k$ be the cluster indicators. There are no cluster-level covariates in our study, but our work can be extended to include them in $V_k$.  We proceed in the counterfactual framework. The counterfactual failure time under treatment $a_j$ for individual $i$ in cluster $k$ is defined as $T_{ik} (a_j)$, $ \forall a_j \in \mathscr{A}$. We similarly define $C_{ik}(a_j)$  as the counterfactual censoring time under treatment $a_j$. Throughout, we maintained the standard assumptions for drawing causal inference with observational clustered survival data: \cite{hu2021estimating,chen2001causal,arpino2016propensity}
\begin{enumerate}
\itemsep-0.3em
\item[(A1)] Consistency: the observed failure time  $T_{ik}=\sum_{j=1}^J T_{ik}(a_j)I(A_{ik} = a_j)$ and censoring time $C_{ik}=\sum_{j=1}^J C_{ik}(a_j)I(A_{ik} = a_j)$, where $I(\cdot)$ is the usual indicator function;
\item[(A2)] Weak unconfoundedness: $T_{ik} (a_j) \amalg A_{ik}  \mid \bm{X}_{ik}, V_k \text{ for } A_{ik}= a_j, a_j \in \mathscr{A}$;
\item[(A3)] Positivity: the generalized propensity score for treatment assignment $e(\bm{X}_{ik}, V_k)=P(A_{ik}=a_j|\bm{X}_{ik},V_k)$ is bounded away from 0 and 1 for $A_{ik}= a_j, a_j \in \mathscr{A}$;
\item[(A4)] Covariate-dependent censoring: 
$T_{ik} (a_j) \amalg C_{ik}(a_j) \mid \bm{X}_{ik}, V_k, A_{ik}$, for $A_{ik}= a_j, a_j \in \mathscr{A}$.
\end{enumerate}
The counterfactual outcomes are linked to the observed outcomes via Assumption (A1). It allows us to write $Y_{ik}=\sum_{j=1}^J Y_{ik}(a_j)I(A_{ik} = a_j)$, where $Y_{ik}(a_j)=\min\lp T_{ik}(a_j),C_{ik}(a_j) \rp$. Similarly, $\Delta_{ik}=\sum_{j=1}^J \Delta_{ik}(a_j)I(A_{ik} = a_j)$, where $\Delta_{ik}(a_j)=I\{T_{ik}(a_j)<C_{ik}(a_j)\}$. Assumption (A2) is referred to as the ``no unmeasured confounding'' assumption.  Because this is an untestable assumption,  we will develop a sensitivity analysis approach in Section~\ref{sec:sa_methods} to gauge the impact of violations of this assumption. Assumption (A3) requires that the treatment assignment is not deterministic within each strata formed by the covariates. \citep{hernan2006estimating} This assumption can be directly assessed by visualizing the distribution of estimated generalized propensity scores. Finally, Assumption (A4) states that the counterfactual survival time is independent of the counterfactual censoring time given pre-treatment covariates, cluster-level variables and treatment variable. This condition directly implies $T_{ik} \amalg C_{ik} \mid \bm{X}_{ik},V_k, A_{ik}$, and is akin to the (conditionally) independent censoring assumption in the traditional survival analysis literature. \citep{hernan2020causal} 

We define the causal estimands directly in terms of counterfactual survival times.  Alternatively, one can define causal estimands based on functionals (e.g., median) of the counterfactual survival curves. \citep{hu2021estimating} In this paper, we focus on the average treatment effect (ATE) defined either over the sample or the population.  Consider a pairwise comparison between treatments $a_j$ and $a_{j'}$. Common sample estimands are the sample average treatment effect (SATE), 
$$\frac{1}{N} \sum_{k=1}^K \sum_{i=1}^{n_k} [ T_{ik}(a_j) - T_{ik}(a_{j'}) ].$$
Common population estimands are the population average treatment effect (PATE), 
$$E[T(a_j) -T(a_{j'})].$$
Conditional average treatment effect (CATE) 
$$\frac{1}{N}\sum_{k=1}^K \sum_{i=1}^{n_k} E \lsq T_{ik}(a_j)-T_{ik}(a_{j'}) \cond \bm{X}_{ik}, V_k \rsq$$ 
is another estimand that preserves some of the properties of the previous two. \citep{hill2011bayesian} As our methods are developed in a Bayesian framework, CATE is a natural estimand to use in this paper. \citep{hill2011bayesian} We obtain the sample marginal effects by averaging the individual conditional expectation of the counterfactual survival times  $E[T_{ik}(a_j)-T_{ik}(a_{j'}) \cond \bm{X}_{ik}, V_k]$ across the empirical distribution of $\{\bm{X}_{ik}, V_k\}_{i=1,k=1}^{n_k,K}$. \citep{hu2020estimation} Another causal estimand of interest is the average treatment effect on the treated (ATT). By averaging the differenced counterfactual survival times over those in the reference group, we can define the ATT counterparts of all three estimands described above. For brevity of exposition,  we focus on CATE in this paper, but our methods can be straightforwardly extended for the ATT effects. For example, the conditional average treatment effect among those who received treatment $a_j$ $\text{CATT}_{a_j \cond a_j, a_{j'}}$ is $$\frac{1}{N_j}\mathop{\sum\sum}_{i,k:A_{ik}=a_j} E \lsq T_{ik}(a_j)-T_{ik}(a_{j'}) \cond \bm{X}_{ik}, V_k \rsq,$$ where $N_j = \sum_{i=1}^{n_k}\sum_{k=1}^K I(A_{ik} =a_j)$ is the size of the reference group $a_j$.

\subsection{The riAFT-BART model for clustered survival data}
For clustered survival data, we propose the following random-intercept AFT model utilizing the likelihood-based machine learning technique BART: 
\begin{eqnarray}
\begin{split}
\label{eq: riAFT-BART}
&\log T_{ik} = f(A_{ik}, \bm{X}_{ik}) + b_k + \epsilon_{ik}, \\
 b_k \distas {i.i.d} & N(0,\alpha_k \tau^2), \quad \epsilon_{ik} \distas {i.i.d} N(0,\sigma^2), \quad b_k \perp \epsilon_{ik}, 
\end{split}
\end{eqnarray} 
where  $f(A_{ik}, \bm{X}_{ik})$ is an unspecified function relating treatments and covariates to survival times $T_{ik}$,  $b_k$'s are the random intercepts for  cluster-specific main effects capturing the institutional variation, and $\epsilon_{ik}$ is the residual term. We use BART to flexibly model the unknown function $f$ by a sum of shallow trees $f(A_{ik}, \bm{X}_{ik})=\sum_{h=1}^H g(A_{ik}, \bm{X}_{ik};\mathcal{W}_h,\mathcal{M}_h)$, where  $\mathcal{W}_h$ is the $h$th binary tree structure,  $\mathcal{M}_h = (\mu_{1h}, \ldots, \mu_{{c_h}h} )^T$ is the set of $c_h$ terminal node parameters associated with tree structure $\mathcal{W}_h$.  For a given value $A_{ik}$ and $\bm{X}_{ik}$ in the predictor space, the assignment function $g(A_{ik}, \bm{X}_{ik};\mathcal{W}_h,\mathcal{M}_h)$ returns the parameter $\mu_{lh}, l \in \{1, \ldots, c_h\}$ associated with the terminal node of the predictor subspace in which $\{A_{ik}, \bm{X}_{ik}\}$ falls. We will place regularizing priors on $\{\mathcal{W}_h,\mathcal{M}_h\}$ to keep the impact of each individual tree on the overall fit small and prevent overfitting. \cite{chipman2010bart,hill2011bayesian,tan2019bayesian} We assume a mean-zero normal distribution for and independence between $b_{k}$ and $\epsilon_{ik}$ with variance $\alpha_k \tau^2$ and $\sigma^2$, respectively. Here we adopt the parameter expansion technique \citep{gelman2008using} and introduce a redundant parameter $\alpha_k$ for the variance of $b_k$ to improve computational performance of our proposed Markov chain Monte Carlo (MCMC)  algorithm (Section~\ref{sec:est_effect}). 

Model~\eqref{eq: riAFT-BART} has two main advantages. First, it allows for direct specification of treatment effect on changes in life expectancy, which substantially facilitates the interpretability of an sensitivity analysis.  Second, unlike the proportional hazards regression, the AFT model formulation can naturally incorporates BART to not only flexibly capture arbitrarily complex functional form of $f(A_{ik},\bm{X}_{ik})$ but also  provide coherent inferences based on a probability model and proper representations of uncertainty intervals via the posterior. 

We decompose the joint prior distribution as 
\begin{align*}
    P\left[\left(\mathcal{W}_{1},\mathcal{M}_{1}\right),\ldots,\left(\mathcal{W}_{H},\mathcal{M}_{H}\right),\sigma,\tau,\alpha_k\right] &= \left[\prod_{h=1}^{H} P\left(\mathcal{W}_{h}, \mathcal{M}_h\right) \right] P(\sigma)P(\tau)P(\alpha_k)\\
    &= \left[\prod_{h=1}^{H} P\left(\mathcal{M}_h \cond \mathcal{W}_{h} \right) P\left(\mathcal{W}_{h}\right)\right] P(\sigma)P(\tau)P(\alpha_k) \\
    &=\left[\prod_{h=1}^{H}\left\{ \prod_{l=1}^{c_{h}}P\left(\mu_{lh}\mid \mathcal{W}_{h}\right)\right\} P\left(\mathcal{W}_{h}\right)\right]P(\sigma)P(\tau)P(\alpha_k).
\end{align*}
Following Chipman et al. \citep{chipman2010bart} and Hendersen et al., \cite{henderson2020individualized} we center the observed responses $y_{ik}$ via the following two steps: (i) fit a parametric intercept-only AFT model assuming log-normal residuals, and estimate the intercept $\hat{\mu}_{AFT}$ and the residual scale $\hat{\sigma}_{AFT}$; (ii) transform the responses as $y_{ik}^{cent} = y_{ik}\exp\lp-\hat{\mu}_{AFT} \rp$. Then for the terminal node values $\mu_{lh}$, we place the prior distribution $\mu_{lh} \sim N \lp 0, \xi^2/(4Hk^2) \rp$, where $k$ is the terminal node $\mu_{lh}$ hyperparameter and $\xi = 4\hat{\sigma}_{AFT}$. This prior induces a $N(0, 4\hat{\sigma}_{AFT}^2/k^2)$ prior on the regression function $f(\cdot)$ in model~\eqref{eq: riAFT-BART}, and with a default setting of $k=2$ assigns 95\% prior probability to the interval $[-2\hat{\sigma}_{AFT}, 2\hat{\sigma}_{AFT}]$, which is sensible.\citep{henderson2020individualized} As suggested in Chipman et al., \citep{chipman2010bart}, we use an inverse gamma distribution $IG(\frac{\nu}{2}, \frac{\nu \lambda}{2})$ as the prior for $\sigma^2$ and the default value for $\nu$ ($\nu =3$), and defer to the defauls for other hyperparameters of the BART trees $\{\mathcal{W}_h, \mathcal{M}_h\}$.  We place a prior $IG(1,1)$ on $\tau^2$ and $\alpha_k$.\citep{gelman2008using} For initial values, we first set an initial random intercept $\hat{b}_k^{(0)}$ to be the mean of the lognormal residuals from a parametric AFT model with $\bm{X}_{ik}$ as the predictors for each cluster $k$. We then set the initial value ${\sigma}^{(0)}$ to be the standard deviation of the pooled model residuals over the $K$ clusters, and initialize $\lambda$ as the value such that $P\left(\sigma < {\sigma}^{(0)}; \nu, \lambda^{(0)}\right) = 0.9$. 
 
We  use data augmentation to deal with right censoring. \citep{henderson2020individualized,roy2017bayesian}  Working with the centered responses $y_{ik}^{cent}$, when $\Delta_{ik} =0$, we impute the unobserved and centered survival times $z_{ik}$ from a truncated normal distribution: 
%$\log z_{ik} \sim$ Truncated-Normal $\lp f(A_{ik}, \bm{X}_{ik})+b_k, \sigma^2; \log y^{cent}_{ik} \rp$: 
$$\lsq \log Z_{ik} \cond \log Z_{ik}>\log y^{cent}_{ik} \rsq \sim N_{(\log y_{ik}^{cent},\infty)}\lp f(A_{ik}, \bm{X}_{ik})+b_k,\sigma^2\rp$$ 
in each Gibbs iteration, where  $\log Z_{ik} \sim N \lp f(A_{ik}, \bm{X}_{ik})+b_k, \sigma^2 \rp$.  The centered complete-data survival times are 
\[
  y^{cent,c}_{ik} = 
  \begin{cases}
    y^{cent}_{ik} \text{ \; \; if } \Delta_{ik} = 1 \\
    z_{ik} \text{ \; \; if } \Delta_{ik} = 0 
  \end{cases}.
\]
\subsection{Posterior inferences for treatment effects} \label{sec:est_effect}
Here we employ a Metropolis within Gibbs procedure for posterior inferences about treatment effects on patient survival. Using the centered complete-data survival times $y_{ik}^{cent,c}$, the joint posterior is 
\begin{align*}
    &\hspace{12pt} P\left(b_k, \tau^2, \alpha_k, \mu_{lh}, \sigma^2 \cond y^{cent,c}_{ik}, \bm{X}_{ik}, A_{ik}, V_k,\{\mathcal{W}_h, \mathcal{M}_h\} \right)\\
    &\propto P \left(y^{cent,c}_{ik} \cond \bm{X}_{ik}, A_{ik}, V_k,b_k, \tau^2, \alpha_k, \sigma^2, \{\mathcal{W}_h, \mathcal{M}_h\} \right) P\left( b_k \cond \tau^2, \alpha_k \right) P\left( \tau^2 \right) P\left( \alpha_k \right) P\lp \mu_{lh}\rp P\left(\sigma^2\right).
\end{align*}
 
We can draw the values of BART sum-of-trees model parameters, $\mu_{lh}$ and $\sigma^2$, directly from the fitted BART model. Their posterior distributions $P \lp \mu_{lh} \cond y^{cent,c}_{ik}, \bm{X}_{ik}, A_{ik}, V_k, b_k, \tau^2, \alpha_k, \sigma^2, \{\mathcal{W}_h\} \rp $ and $P\lp \sigma^2 \cond  y^{cent,c}_{ik}, \bm{X}_{ik}, A_{ik}, V_k, b_k, \tau^2, \alpha_k, \{\mathcal{W}_h, \mathcal{M}_h\} \rp $  are presented in Web Section S1.  We can show  the posterior distribution of the random intercept $b_k$ is 
 $$\lsq b_k \cond y^{cent,c}_{ik}, \bm{X}_{ik}, A_{ik}, V_k, \tau^2, \alpha_k, \sigma^2, \{\mathcal{W}_h, \mathcal{M}_h\}\rsq \sim N\lp \dfrac{\tau^2 \alpha_k \sum_{i=1}^{n_k}\lp y^{cent,c}_{ik}- \hat{f}(\bm{X}_{ik}, A_{ik})\rp}{n_k\tau^2\alpha_k+\sigma^2},  \dfrac{\sigma^2 \tau^2 \alpha_k}{n_k \tau^2\alpha_k+\sigma^2} \rp.$$
 The posterior of $\alpha_k$,  used for parameter expansion, is 
 $$\lsq \alpha_k \cond y^{cent,c}_{ik}, \bm{X}_{ik}, A_{ik}, V_k, \tau^2, b_k, \sigma^2, \{\mathcal{W}_h, \mathcal{M}_h\}\rsq \sim IG \lp 1, 1+ \frac{\sum_{k=1}^K b_k^2}{2\tau^2}\rp.$$
We obtain the posterior of $\tau^2$ as 
$$\lsq \tau^2 \cond y^{cent,c}_{ik}, \bm{X}_{ik}, A_{ik}, V_k, b_k,\alpha_k, \sigma^2, \{\mathcal{W}_h, \mathcal{M}_h\} \rsq \sim IG \lp \dfrac{K}{2}+1, \dfrac{\sum_{k=1}^K b_k^2+2\alpha_k}{2\alpha_k}\rp.$$ Complete derivation of the posterior distributions  
are provided in Web Section S1. 

We now describe our Metropolis within Gibbs procedure to draw from the posterior distribution of our proposed riAFT-BART model~\eqref{eq: riAFT-BART}. A \emph{single iteration} of our sampling algorithm proceeds through the following steps: 
\vspace{-.5cm}
\begin{algorithm}
\caption{A single iteration of riAFT-BART sampling algorithm} \label{alg:mod-sampling}
\begin{enumerate}
\item  Update $b_k$, $\tau^2$ and $\alpha_k$ from their respective posterior distributions. \item Using $\log y^{cent,c}_{ik} - b_k$ as the responses and $\{A_{ik}, \bm{X}_{ik}\}$ as the covariates, update BART sum-of-trees  model via  parameters $\mu_{lh}$ and $\sigma^2$, using the Bayesian backfitting approach of Chipman et al. \citep{chipman2010bart} Directly update $f(A_{ik}, \bm{X}_{ik})$ using the updated BART model,  for $i = 1, \ldots, n_k, k=1, \ldots, K$. 
\item For each $\{i,k\} \in \{i =1, \ldots, n_k, k =1, \ldots, K\}$, update $z_{ik}$ by sampling 
$$\log z_{ik} \sim \text{Truncated-Normal} \lp f(A_{ik}, \bm{X}_{ik})+b_k, \sigma^2; \log y^{cent}_{ik} \rp.$$
\end{enumerate}
\end{algorithm}

Because we use the centered responses $\log(y^{cent}_{ik}) = \log(y_{ik}) - \hat{\mu}_{AFT}$ in posterior computation, we add $\hat{\mu}_{AFT}$ back to the posterior draws of $f(A_{ik}, \bm{X}_{ik})$ in the final output.

To draw posterior inferences about the CATE effects via riAFT-BART, we note that under the causal assumptions (A1)--(A4), 
\begin{align}
\begin{split}
    &\hspace{13pt} E\lbc \log T_{ik}(a_j) - \log T_{ik}(a_{j'})\rbc \\
    &=E_{\bm{x}_{ik},b_k}\lbc E \lp \log T_{ik} \cond A_{ik}=a_j, \bm{X}_{ik}  = \bm{x}_{ik},b_k \rp -E \lp \log T_{ik} \cond A_{ik}=a_{j'}, \bm{X}_{ik} = \bm{x}_{ik},b_k \rp \rbc\\
    &=E_{\bm{x}_{ik},b_k} \lsq  E\lbc \lp f\lp a_j, \bm{x}_{ik} \rp +b_k +\epsilon_{ik}\rp  -  \lp f\lp a_{j'}, \bm{x}_{ik} \rp +b_k +\epsilon_{ik}\rp \rbc\rsq\\
    &= E_{\bm{x}_{ik}}\lsq E\lbc f\lp a_j, \bm{x}_{ik}\rp - f\lp a_{j'}, \bm{x}_{ik}\rp \rbc \rsq. 
\end{split}
\end{align}
  This allows us to estimate treatment effect via outcome modeling. Specifically,  
  \begin{align}
      \widehat{CATE}_{a_j,a_{j'}} =  \frac{1}{D} \sum_{d=1}^D \frac{1}{n_kK} \sum_{k=1}^K \sum_{i=1}^{n_k} \lbc f^d \lp a_j, \bm{x}_{ik}\rp - f^d \lp a_{j'},\bm{x}_{ik} \rp \rbc,
  \end{align}
  where $f^d$ is the $d^{th}$ draw from the posterior distribution of $f$. Inferences can be obtained based on the $D$ posterior average treatment effects, $(1/n_kK) \sum_{k=1}^K\sum_{i=1}^{n_k}\lbc f^d \lp a_j, \bm{x}_{ik} \rp - f^d \lp a_{j'}, \bm{x}_{ik} \rp \rbc, d =1,\ldots, D$.
 
\section{Sensitivity analysis} \label{sec:sa_methods}
\subsection{Overview}
 The estimation of causal effects with observational data relies on the weak unconfoundedness assumption (A2), which cannot be verified empirically. Violations of this assumption can lead to biased treatment effect estimates.  One widely recognized way to address concerns about violations of this assumption is  sensitivity analysis.  In fact, the STROBE guidelines recommend observational studies be accompanied by sensitivity analysis investigating the ramifications of potential unmeasured confounding. \citep{von2007strengthening}  Many sensitivity analysis methods have been developed, including Rosenbaum's $\Gamma$, \citep{rosenbaum2002covariance}  external adjustment, \citep{kasza2017assessing}  confounding functions \citep{robins1999association,hu2022flexible} and the E value, \citep{vanderweele2017sensitivity} to name a few. These methods differ in how unmeasured confounding is formulated and parameterized. Sensitivity analysis approaches in the context of  multiple treatments and clustered censored survival outcomes are an underdeveloped area. 
 
 With multilevel data, there can be unobserved confounders at both cluster- and individual-level. It has been shown in the literature that with propensity score based methods, the fixed-effects model for propensity score estimation automatically controls for the effects of unmeasured cluster-level confounders. \citep{arpino2011specification,li2013propensity,fuentes2021causal} In situations where the cluster sizes are small,  a random-effects propensity score model may provide more accurate effect estimates, but is reliant on inclusion of important cluster-level covariates as regressors. As the cluster size increases, results from the random-effects model converge to those from a corresponding fixed-effects model. Although there is sparse literature on whether accounting for the clustered structure in potential outcome models would protect against misspecification due to cluster-level confounders,  the outcome model is connected to the propensity score model in that the sufficient statistics (treatment group means of covariates) that must be balanced to eliminate confounding differences under both models are the same. \citep{li2013propensity} Li et al. \citep{li2013propensity} conducted a simulation to show that ignoring the clustered data structure in both the propensity score and outcome models would lead to biased ATE estimates; respecting the structure in at least one of the models gives consistent estimates. Based on these grounds and that we are dealing with large clusters of a national database, we believe our riAFT-BART model~\eqref{eq: riAFT-BART} -- which will converge to a fixed-effects model with large cluster sizes -- will represent heterogeneity in cluster-level unmeasured confounding by the random effects $\{b_k\}$. We then assume by conditioning on $\{b_k\}$ and $\{\bm{X}_{ik}\}$, the potential outcome and treatment at cluster-level are independent, and introduce our sensitivity analysis approach for individual-level unmeasured confounding. 

Our sensitivity analysis approach is along the line of work by Hu et al., \citep{hu2022flexible} and is based within the framework of confounding function. \citep{robins1999association,brumback2004sensitivity} The confounding function based methods have the advantage of avoiding introducing a hypothetical unmeasured confounder and making an assumption about its underlying structure, on which there is a lack of consensus, and are preferred when the primary interest is in understanding the total effect of all unmeasured confounders.\citep{brumback2004sensitivity,hu2022flexible}

\subsection{Confounding function adjusted treatment effect estimates}
For notational brevity, we suppress the $ik$ subscript denoting individual.  Following Brumback et al.\citep{brumback2004sensitivity} and Hu et al., \citep{hu2022flexible} we first define the confounding function for any pair of treatments $(a_j, a_{j'})$ as
\begin{eqnarray} \label{eq:cf}
c(a_j, a_{j'}, \bm{x}, v) = E \lsq \log T(a_j) \cond A = a_j, \bm{X}=\bm{x}, V=v\rsq - E \lsq \log T (a_{j}) \cond A =a_{j'}, \bm{X}= \bm{x}, V=v \rsq.
\end{eqnarray} 
This  confounding function directly represents the difference in the mean potential log survival times under treatment $a_j$ between those treated with $a_j$ and those treated with $a_{j'}$, who have the same level of $\bm{x}$.  Under the assumption of no unmeasured confounding, given measured individual- and cluster-level covariates $\bm{X}$ and $V$, the potential outcome is independent of treatment assignment.  Had they received the same treatment $a_j$, their mean potential survival times would have been the same, or $c(a_j, a_{j'}, \bm{x}, v) = 0$, $\forall \{a_j, a_{j'}\} \in \mathscr{A}$. When this assumption is violated and there exists unmeasured confounding, the causal effect estimates using measured confounders will be biased. The bias in the estimated treatment effect $\widehat{CATE}_{a_j, a_{j'}}$ takes the following form: 
\begin{align} \label{eq:biasform}
\begin{split}
\text{Bias}(a_j,a_{j'} \cond \bm{x},v) =&-p_{j} c(a_{j'}, a_j, \bm{x},v) + p_{j'}c(a_j,a_{j'},\bm{x},v)\\
&-\sum\limits_{m: a_m \in \mathscr{A}\setminus\{a_j, a_{j'}\}} p_{m} \lbc c(a_{j'}, a_m, \bm{x},v) -c(a_j,a_m,\bm{x},v) \rbc, 
\end{split}
\end{align}
where $p_{j} = P(A= a_j \cond \bm{X}= \bm{x}, V=v)$, $j \neq j' \in \{1, \ldots, J\} $.
A proof of this result is presented in Web Section S2. 

Given known confounding functions $c$, we can construct the confounding function adjusted estimators by first modifying the actual survival time $T$, and then estimating the causal effect by fitting our riAFT-BART model to modified outcomes. In this way, the bias in equation~\eqref{eq:biasform} will be effectively removed from the adjusted effect estimate. Because the survival times $T$ may be right censored and we deal with right censoring using data augmentation, the outcome modification can be implemented on the complete-data survival times. We propose a Monte Carlo sensitivity analysis approach along the line of work by Hu et al., \cite{hu2022flexible} which was developed for binary outcomes. We extend their work to accommodate multilevel censored survival outcomes.  Our sensitivity analysis proceeds with steps listed in Algorithm~\ref{alg:SA}. 

\begin{algorithm}
\caption{Sensitivity analysis algorithm} \label{alg:SA}
\begin{enumerate}
    \item Fit a multinomial probit BART model $f^{\text{MBART}}(A_{ik}\cond \bm{X}_{ik}, V_k)$ to estimate the generalized propensity scores, $p_j \equiv P (A_{ik} =a_j \cond \bm{X}_{ik} =\bm{x}_{ik}, V_k=v_k) \;  \forall a_j \in \mathscr{A}$, for each individual. 
    \item For each treatment $a_j \in \mathscr{A}$, draw $Q_1$ generalized propensity scores $\tilde{p}_{m1}, \ldots, \tilde{p}_{mQ_1}, \forall m \neq j \wedge a_m \in \mathscr{A}$ from the posterior predictive distribution of $f^{\text{MBART}}(A_{ik}\cond \bm{X}_{ik}, V_k)$  for each individual. 
    \item For $q \in \{1, \ldots, Q_1\}$, draw $Q_2$ values $\gamma^*_{mq1}, \ldots, \gamma^*_{mqQ_2}$ from the prior distribution of each of the confounding functions $c(a_j, a_m, \bm{x},v)$, for each $m \neq j \wedge a_m \in \mathscr{A}$.
    \item For each treatment $a_j$, adjust the centered complete-data survival times (Step 2 of Algorithm~\ref{alg:mod-sampling}) as follows: 
    \begin{eqnarray}\label{eq:ycf}
     \log y_{ik}^{CF} \equiv \log y_{ik}^{cent,c} - \sum_{m \neq j}^J P\lp A_{ik} = a_m\cond \bm{X}_{ik}= \bm{x}, V_k = v\rp c(a_j, a_m, \bm{x}, v),
    \end{eqnarray}
    for each of $Q_1Q_2$ draws of $\{\tilde{p}_{m1},  \gamma^*_{m11}, \ldots, \gamma^*_{m1Q_2}, \ldots,  \tilde{p}_{mQ_1},  \gamma^*_{mQ_11}, \ldots,  \gamma^*_{mQ_1Q_2};  m \neq j \wedge a_m \in \mathscr{A}\}$. 
    \item Run Algorithm~\ref{alg:mod-sampling} for riAFT-BART on each of $Q_1\times Q_2$ sets of observed data with  $\log y_{ik}^{CF}$. Estimate the combined adjusted causal effects and uncertainty intervals by pooling posterior samples across model fits arising from the $Q_1 \times Q_2$ data sets.  
\end{enumerate}
\end{algorithm}

Note that in Step 1 of Algorithm~\ref{alg:SA},  we can fit a flexible fixed-effects multinomial probit BART model for generalized propensity scores. Steps 2 and 3 constitute a nested multiple imputation, \citep{rubin2003nested} which is used to draw samples for the product term $P\lp A_{ik} = a_m \cond \bm{X}_{ik}=\bm{x}, V_k=v\rp c(a_j, a_m,\bm{x},v)$ in equation~\eqref{eq:ycf}. Step 4 ``corrects" the complete-data survival times to adjust the treatment effect estimate for individual-level unmeasured confounding.  This is because the causal effect is defined as the between-group difference in mean potential outcomes and is estimated based on the observed outcomes.  To correct the bias in equation~\eqref{eq:biasform} due to individual-level unmeasured confounding, we adjust the actual survival time $T$ of an individual who received treatment $a_j$ as 
\begin{eqnarray*}
\log T^{CF} &=& \log T - \lsq E \lbc \log T(a_j) \cond a_j, \bm{x}, v\rbc -  E \lbc  \log T(a_j) \cond \bm{x}, v \rbc \rsq. 
\end{eqnarray*} 
Because the survival time $T$ can be right censored, we can replace the centered complete-data survival time $\log y_{ik}^{cent,c}$ used in our riAFT-BART sampling Algorithm~\ref{alg:mod-sampling} for treatment effect estimation with adjusted $\log y_{ik}^{CF}$ as in equation~\eqref{eq:ycf}. Web Section S2 provides a detailed justification of this strategy for obtaining confounding function adjusted causal effect estimates.  
Finally, we obtain the overall estimates of the adjusted causal effect  and sampling variance  as the posterior mean and variance of the pooled posterior samples  across the $Q_1 \times Q_2$ data sets. \citep{zhou2010note} We used $Q_1=30$ and $Q_2=30$ when implementing our sensitivity analysis Algorithm~\ref{alg:SA} in both simulation (Web Section S3) and case study (Section~\ref{sec:application}). 

The confounding functions are not identifiable from the observe data. We can assume the form of confounding function to represent our prior beliefs about the possible direction and magnitude of the effect of unmeasured confounding. We follow strategies discussed in earlier work \citep{robins1999association, brumback2004sensitivity, hogan2014bayesian, hu2022flexible} to specify the signs and bounds of the confounding functions. For example, by assigning $c(a_j, a_{j'}, \cond \bm{x},v) >0$ and  $c(a_{j'}, a_j, \cond \bm{x},v) < 0$, we assume the unmeasured factors tend to lead clinicians to systematically prescribe $a_j$ to healthier patients relative to $a_{j'}$, because  patients treated with $a_j$ will on average have longer potential survival time to both $a_j$ and $a_{j'}$  than patients treated with $a_{j'}$. Web Table 1 presents interpretations of confounding functions with other specifications of signs. When setting the bounds, we assume the unmeasured confounding would account for less than $\omega$ units of the remaining standard deviation unexplained by measured confounders $\bm{X}_{ik}$. Using the NCDB data in our case study as an example, a specification of $c(1,2 \cond \bm{x},v) \leq \hat{\sigma} = 0.90$ and $c(2,1 \cond \bm{x},v) \geq -\hat{\sigma} = -0.90$  assumes that patients assigned to RP will on average have $\exp(0.90) =2.45$ months longer potential survival times than patients assigned to EBRT+AD to both treatment options; and therefore clinicians tend to prescribe RP to healtier patients. This bound of $\omega=1$ unit of remaining standard deviation is a plausible assumption. In the NCDB data, the median survival time was 7.7 and 7.8 years for EBRT+AD and RP group respectively and was not reached for the EBRT+brachy$\pm$AD treatment group (Web Figure 1). As suggested by Hu et al.\citep{hu2022flexible} we draw the values of confounding functions from the uniform distribution. 

\section{Simulation} \label{sec:sim}
\subsection{Comparison methods}
Through a contextualized simulation, we investigate the practical operating characteristics of our proposed method riAFT-BART. We also adapt the popularly used inverse probability weighting method into the setting of clustered and censored survival data to form two comparison methods: inverse probability of treatment weighting with the random-intercept Cox regression model (IPW-riCox) and  doubly robust random-intercept additive hazards model (DR-riAH). In addition, we consider another outcome modeling based method, random-intercept generalized additive proportional hazards model (riGAPH), \citep{hastie1990generalized, hu2021estimating} that is flexible at capturing nonlinear relationships.  We use the counterfactual survival curve as the basis to objectively compare methods. Note that we can derive the individual survival curve corresponding to our riAFT-BART model~\eqref{eq: riAFT-BART} as 
 \begin{eqnarray}
 \label{eq:indiv-cf-surv}
 P \lp T_{ik}>t \cond A_{ik}, \bm{X}_{ik}, \sigma, b_{k}\rp = 1- \Phi \lp \frac{\log t -f(A_{ik}, \bm{X}_{ik}) - b_k}{\sigma}\rp.
  \end{eqnarray}
We can define the causal estimands by contrasting the conditional survival probability up to a fixed time $t^*$, or by comparing the conditional restricted mean survival time (RMST). \citep{royston2013restricted} For some arbitrary time bound $t^*$, the RMST can be represented as the area under the survival curve $S(t)$ up to $t^*$, $\text{RMST} = \int_0^{t^*} S(t) dt$. In our simulation, we present results based on both metrics, because  in our motivating prostate cancer research question, 5-year survival and RMST are of most clinical relevance. \citep{ennis2018brachytherapy}

For the DR-riAH method, following  suggestions by Li et al.,\citep{li2013propensity} we obtain the DR-riAH treatment effect estimator based on the survival probability at $t^*$ as
\begin{equation*}
\widehat{CATE}_{a_j,a_{j'}} = \frac{1}{n_kK} \sum_{k=1}^K \sum_{i=1}^{n_k} \lp\frac{A_{ik}
P(T_{ik}> t^*)-(A_{ik}-\hat{e}_{ik})P\lp T_{ik}(a_j) >t^* \rp}{\hat{e}_{ik}} - \frac{(1-A_{ik})
P(T_{ik}> t^*)-(A_{ik}+\hat{e}_{ik})P\lp T_{ik}(a_{j'}) >t^* \rp}{1-\hat{e}_{ik}}  \rp,
\end{equation*}
where $\hat{e}_{ik} = P\lp A_{ik} = a_m\cond \bm{X}_{ik}= \bm{x}, V_k = v\rp$ is the estimated generalized propensity score, the observed survival probability $P(T_{ik} >t^*)$ is the Kaplan-Meier estimator and the predicted counterfactual survival probability $P\lp T_{ik}(a_{j'}) >t^* \rp$ is calculated from the random-intercept additive hazards model. \citep{cai2011additive} For CATE effects based on RMST, we replace the observed (counterfactual) survival probability with the area under the observed (counterfactual) survival curve. 

To  assess the performance of each method, we compare the relative bias defined as  $$\frac{\widehat{CATE}_{a_j, a_{j'}} - CATE^0_{a_j,a_{j'}}}{CATE^0_{a_j,a_{j'}}},$$ 
where $CATE^0_{a_j,a_{j'}}$ is the true treatment effect, and the frequentist coverage probability for $\widehat{CATE}_{a_j, a_{j'}}$ in terms of 5-year survival probability and 5-year RMST among 250 data replications. 

When implementing the comparison methods, for weighting based methods IPW-riCox and DR-riAH, we used Super Leaner\citep{van2007super}  to estimate the stabilized inverse probability of treatment weights for improved modeling flexibility and accuracy of the estimated weights. Super Learner was implemented via the $\R$ package $\code{SuperLearner}$ with \code{SL.library = c("SL.xgboost", "SL.bartMachine","SL.gbm")}.  We fitted a weighted random-intercept Cox regression model using the \code{coxme} function of $\R$ package \code{coxme} to obtain the IPW-riCox estimator. For DR-riAH, we fitted the random-intercept additive hazards model using the \code{aalen} function from $\R$ package \code{timereg} to compute the counterfactual outcomes used in the DR-riAH estimator.  To implement ri-GAPH, we used the \code{gam} function from $\R$ package \code{mgcv} and  two helper functions (\code{as\_ped} and \code{add\_surv\_prob}) from $\R$ package \code{pammtools}. RMST was calculated by the trapezoidal rule using the $\R$ function \code{rmst} of \code{RISCA} package. For all methods, the same confounders available to the analyst were used in the linear forms in the corresponding models.

\subsection{Simulation design}
Our data generating processes are contextualized in NCDB data settings.  We generate $K=20$ clusters, each with a sample size of $n_k=500$, and the total sample size is $N=10000$. We simulate 10 confounding variables, with five continuous variables independently generated from the standard normal distribution $X_{ikj} \sim N(0,1), \; j=1, \ldots, 5$, two categorical variables independently generated from the multinomial distribution $X_{ikj} \sim \text{Multinomial}(1,.3,.3,.4), \; j=6, 7$ and three binary variables $X_{ik8} \sim \text{Bern} (0.6)$,   $X_{ik9} \sim \text{Bern} (0.4)$,  $X_{ik10} \sim \text{Bern} (0.5)$ generated for each individual $i$ in cluster $k$. Throughout we consider three treatment groups. The treatment assignment mechanism follows a random intercept multinomial logistic regression model, 

\begin{equation}\label{eq:trt_assign}
\begin{split}
\ln  \dfrac{P(A_{ik}=1)}{P(A_{ik}=3)} &= \xi_{01} + \bm{X}_{ik}\xi_1^L + \bm{G}_{ik}\xi_1^{NL} + \tau_k \\
\ln  \dfrac{P(A_{ik}=2)}{P(A_{ik}=3)} &= \xi_{02} + \bm{X}_{ik}\xi_2^L + \bm{G}_{ik}\xi_2^{NL} + \tau_k \\
\end{split}
\end{equation}
where $\tau_k \sim N(0,1^2)$,  $\bm{G}_{ik}$ denotes the nonlinear transformations and higher-order terms of the predictors $\bm{X}_{ik}$, and $\xi^{L}_{1}, \xi^{L}_{2}$  and $\xi^{NL}_{1},\xi^{NL}_{2}$ are respectively vectors of coefficients for the untransformed versions of the confounders $\bm{X}_{ik}$ and for the transformed versions of the confounders captured in $\bm{G}_{ik}$. The intercepts $\xi_{01}$ and $\xi_{02}$ control the ratio of units across three treatment groups, for which we use 6:3:1 to mimic the ratio of individuals in the NCDB data. 

We generate the potential survival times from a Weibull survival curve, 
\begin{eqnarray}\label{eq:Smod}
 S_{ik}(t) = \exp \lsq -\lbc\lambda_{a_j}\exp\lp \beta^L_{a_j},\beta^{NL}_{a_j};\bm{X}_{ik},\bm{G}_{ik},b_k \rp t\rbc ^\eta \rsq,
\end{eqnarray}
where  $\beta^L_{a_j}$ is a treatment-specific vector of coefficients for $\bm{X}_{ik}$ and $\beta^{NL}_{a_j}$ for $\bm{G}_{ik}$, $\forall a_j \in  \{1,2,3\}$. 
We explicitly avoid generating survival times from a lognormal AFT model to assess the robustness of the assumption of lognormal residuals in our riAFT-BART model formulation in equation~\eqref{eq: riAFT-BART}. The parameter $\eta$ is set to 2 and $\exp(0.7+0.5x_1)$ to respectively produce proportional hazards (PH) and nonproportional hazards (nPH). Three sets of non-parallel response surfaces are generated as 
\begin{eqnarray}\label{eq:Tmod}
 T_{ik}\lp a_j \rp = \lbc \frac{-\log U}{\lambda_{a_j}\exp\lp \bm{X}_{ik}\beta^L_{a_j} + \bm{G}_{ik}\beta^{NL}_{a_j}+b_k\rp} \rbc ^{1/\eta}
\end{eqnarray}
for $a_j \in \{1,2,3\}$, where $U$ is a random variable following the uniform distribution on the interval $[0,1]$, $b_k \sim N(0,4^2), \lambda_{a_j} = \{3000, 1200, 2000\}$ for  
 $a_j = 1,2,3$.  Observed and uncensored survival times are generated as $T_{ik} = \sum_{a_j \in \{1,2,3\}} T_{ik}(a_j)I(A_{ik}= a_j)$. We further generate censoring time $C$ independently from an exponential distribution with the rate parameter selected to induce two different censoring proportions: 10\% and 40\%.

 The data generating processes will produce four configurations: (PH vs. nPH) $\times$ (10\% censoring proportion vs. 40\% censoring proportion).  Detailed model specifications for treatment assignment, as in model~\eqref{eq:trt_assign}, and for outcomes, as in model~\eqref{eq:Tmod}, are given in Web Table 2. Web Figure 2 presents the Kaplan-Meier survival curves stratified by treatment for each of four data configurations in our simulation. The assessment of covariate overlap displayed in Web Figure 3 suggests there is moderate to strong overlap across three simulated treatment groups, which represents the overlap in the NCDB dataset (Web Figure 4).   
 
 We conduct an illustrative simulation, using one individual-level binary measured confounder and one individual-level binary unmeasured confounder, to examine how our sensitivity analysis approach performs in comparison to two causal analyses: (i) including unmeasured confounders and (ii) ignoring unmeasured confounders. Simulation design details are provided in Web Section S3. 
 
\begin{figure}[ht]
\centering
\includegraphics[width = .9\textwidth]{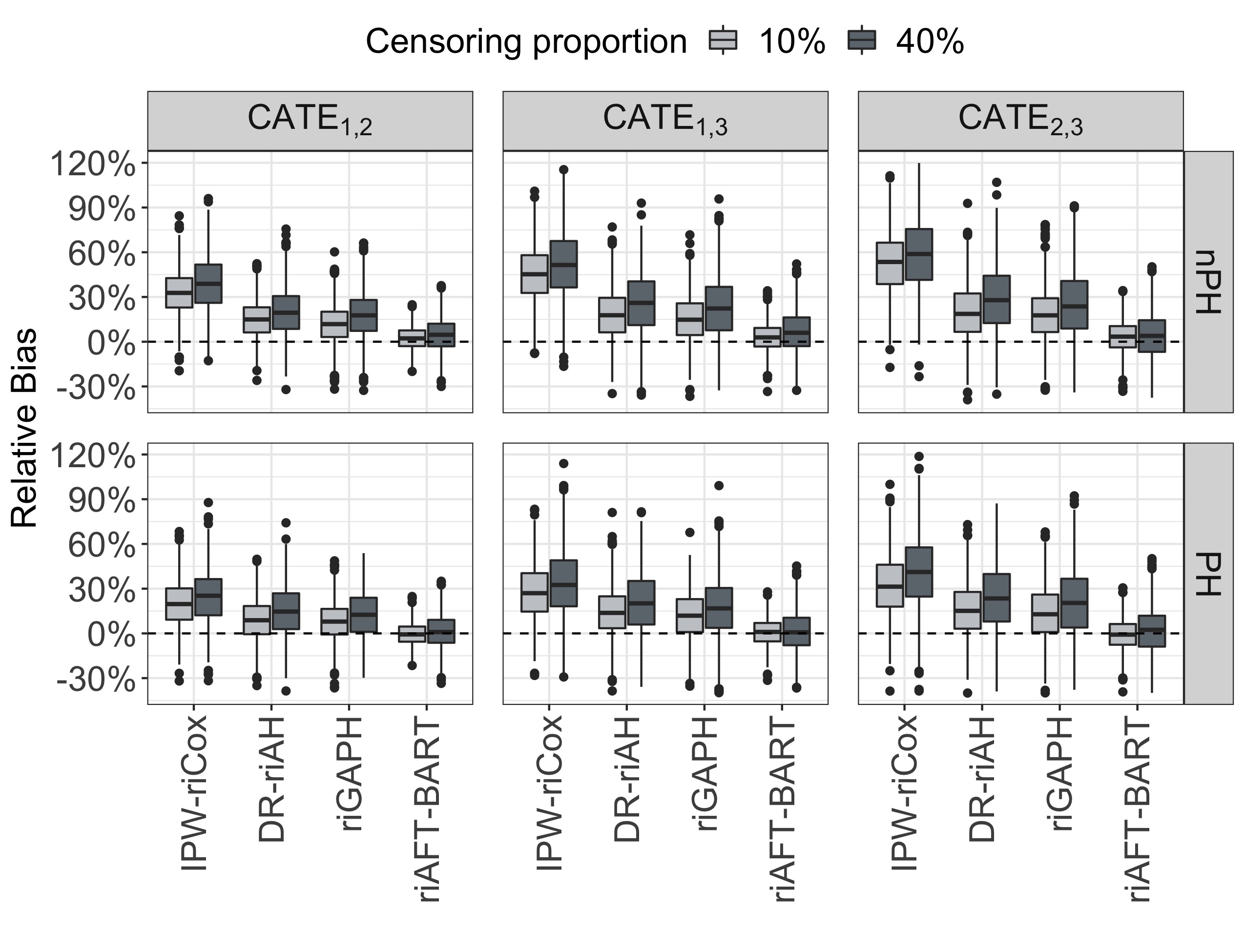}
\caption{Relative biases among 250 replications for each of four methods, IPW-riCox, DR-riAH, riGAPH and riAFT-BART, and three treatment effects  $CATE_{1,2}$, $CATE_{1,3}$ and $CATE_{2,3}$ based on 5-year RMST under four data configurations: (proportional hazards vs. nonproporitonal hazards) $\times$ (10\% censoring proportion vs. 40\% censoring proportion). The true treatment effects under proportional hazards are $CATE^{0,PH}_{1,2} = 7.7$ months,  $CATE^{0,PH}_{1,3} = 3.6$  months and $CATE^{0,PH}_{2,3} = -4.1$ months. The true treatment effects under nonproportional hazards are $CATE^{0,nPH}_{1,2} = 8.1$ months, $CATE^{0,nPH}_{1,3} = 3.9$ months and $CATE^{0,nPH}_{2,3} = -4.2$ months.}
\label{fig:bias-rmst}
\end{figure}
  
\begin{figure}[ht]
\centering
\includegraphics[width = .8\textwidth]{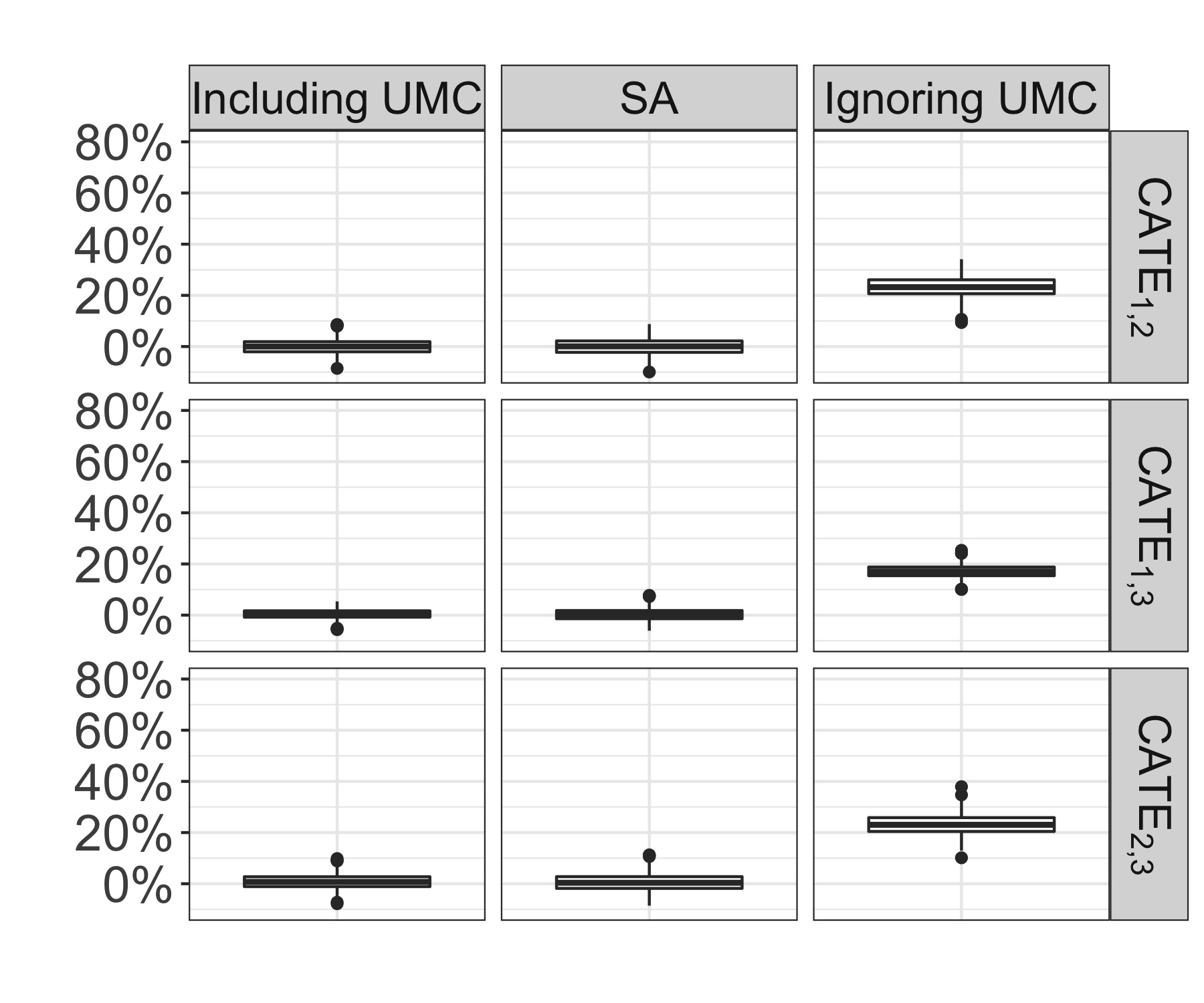}
\caption{Relative biases in the estimates of three pairwise treatment effects $\text{CATE}_{1,2}$, $\text{CATE}_{1,3}$ and $\text{CATE}_{2,3}$ among 1000 replications using data simulated for illustrative sensitivity analysis. Three causal analyses were performed: (i)  including unmeasured confounder (Including UMC), (ii) sensitivity analysis given known confounding functions (SA), and (iii) ignoring unmeasured confounder (Ignoring UMC).}
\label{fig:sim-SA}
\end{figure}

\subsection{Results}
Figure~\ref{fig:bias-rmst} displays boxplots of relative biases in three treatment effect estimates $\widehat{CATE}_{1,2}$, $\widehat{CATE}_{1,3}$ and $\widehat{CATE}_{2,3}$ based on 5-year RMST, among 250 simulations under four data configurations. Our proposed method riAFT-BART boasts the smallest biases and variability in all treatment effect estimates across all simulation scenarios, followed by DR-riAH and riGAPH; while IPW-riCox yields the largest biases and variability.  When the censoring proportion increases, all four methods show decreased performance. The violation of proportional hazards has the largest impact on the IPW-riCox method, demonstrated by the largest bias increase in CATE estiamtes, but only a small impact on riAFT-BART. Even though our method does not require proportional hazards, the elevated data complexity in the non-proportional hazards setting -- the shape parameter in the outcome model~\eqref{eq:Tmod} is covariate-dependent --  may have contributed to  
the slight increase in the biases. The bias assessment based on 5-year survival probability is provided in Web Figure 5, which conveys the same messages as the RMST based results.

\begin{table}[ht]
\centering
\caption{The coverage probability for three treatment effect estimates $\widehat{CATE}_{1,2}$, $\widehat{CATE}_{1,3}$ and $\widehat{CATE}_{2,3}$ based on 5-year RMST under four data configurations: (proportional hazards vs. nonproporitonal hazards) $\times$ (10\% censoring proportion vs. 40\% censoring proportion).} 
\begin{tabular}{clccccccc}
\toprule
& &  \multicolumn{3}{c}{Proportional hazards} && \multicolumn{3}{c}{Nonproportional hazards}  \\
\cmidrule{3-5}  \cmidrule{7-9}
 Censoring \%  &Methods & $\text{CATE}_{1,2}$ & $\text{CATE}_{1,3}$ & $\text{CATE}_{2,3}$ & & $\text{CATE}_{1,2}$ & $\text{CATE}_{1,3}$ & $\text{CATE}_{2,3}$ \\
\midrule
\multirow{4}{*}{10\%}  & IPW-riCox &25.2  &29.2  & 28.8 & & 17.2 & 22.4 &19.6\\
& DR-riAH & 81.6 &81.2 &82.0 && 79.2 & 79.6 & 79.6\\
&ri-GAPH & 84.4	&85.2 & 84.8&& 82.4	& 84.0 &  83.2\\
&riAFT-BART & 	94.8 &	95.2 & 95.2   && 94.4 &	94.4 & 94.8\\
\midrule
\multirow{4}{*}{40\%}  &IPW-riCox &22.0 &26.4 & 24.0 && 13.2&19.6 &16.8\\
& DR-riAH  &78.4 &79.6	& 79.6  && 76.2	&78.8 & 77.2 \\
& ri-GAPH   &81.2 &83.2	& 82.0 &&79.2 &82.0	 & 80.4  \\
& riAFT-BART &94.8 &95.2 & 94.8 && 94.0	& 93.2 & 94.0 \\
\bottomrule
\end{tabular}
\label{tab:CP}
\end{table}

Table~\ref{tab:CP} presents the frequentist coverage probability of each of four estiamtors for each simulation configuration. The proposed method riAFT-BART provides nominal frequentist coverage probability under proportional hazards with 10\% censoring proportion. Even in the most complex data settings with nonproportional hazards and 40\% censoring proportion, riAFT-BART still provides close-to-nominal frequetist coverage probability. By comparison, 
the IPW-riCox estimator is the least efficient producing unsatisfactory coverage probability; DR-riAH and ri-GAPH deliver similar coverage probabilities around 0.8.  Web Table 3 examines the frequentist coverage probability of the estimators based on 5-year survival probability, and we observed the same differences across the estimators demonstrated in the RMST based results. 

Web Figure 6 suggests that our riAFT-BART algorithm converges well by plotting 3500 posterior draws of the variance parameters $\tau$ and $\sigma$, and cluster-specific parameter $\alpha_k$ and the random intercepts $b_k$ for clusters $k=1$, $k=10$ and $k=20$. 

The illustrative simulation for sensitivity analysis, displayed in Figure~\ref{fig:sim-SA}, empirically supports our proposed sensitivity analysis Algorithm~\ref{alg:SA}. Given the known confounding functions, our sensitivity analysis estimators are similar to the results that could be achieved had the unmeasured confounders been made available to the analyst. The naive analysis where we ignored the unmeasured confounding produced substantially biased estimators.

\section{Application to prostate cancer} \label{sec:application}
We applied the proposed method riAFT-BART to estimate the comparative effectiveness of three treatment approaches, RP, EBRT+AD and EBRT+brachy$\pm$AD, on patient survival among high-risk localized prostate cancer patients who are older than 65 years of age. We then applied the proposed sensitivity analysis approach to evaluate how the causal conclusions about treatment effects would change in response to various degrees of departure from the no unmeasured confounding assumption. 

The analysis dataset was drawn from the NCDB and included 23058 high-risk localized prostate cancer patients who were older than 65 when diagnosed between 2004 and 2015.  Among these patients, 14237 received RP, 6683 undertook EBRT+AD with at least 7920 cGy EBRT dose \citep{hu2021estimating} and 2138 were treated with  EBRT+brachy$\pm$AD. Included in the dataset are pre-treatment patient information on age, race and ethnicity, insurance status, income, education level, clinical T stage, year of diagnosis, prostate-specific antigen  and gleason score, and geographic locations of treating facilities. There are nine hospital locations, which were considered as the clusters in our analysis.  Detailed descriptions of the individual- and cluster-level variables (hospital locations) are presented in Web Table 4. Covariates are deemed to have good overlap across three treatment groups based on the estimated generalized propensity scores shown in Web Figure 4. 

\begin{figure}[ht]
\centering 
\includegraphics[width = .8\textwidth]{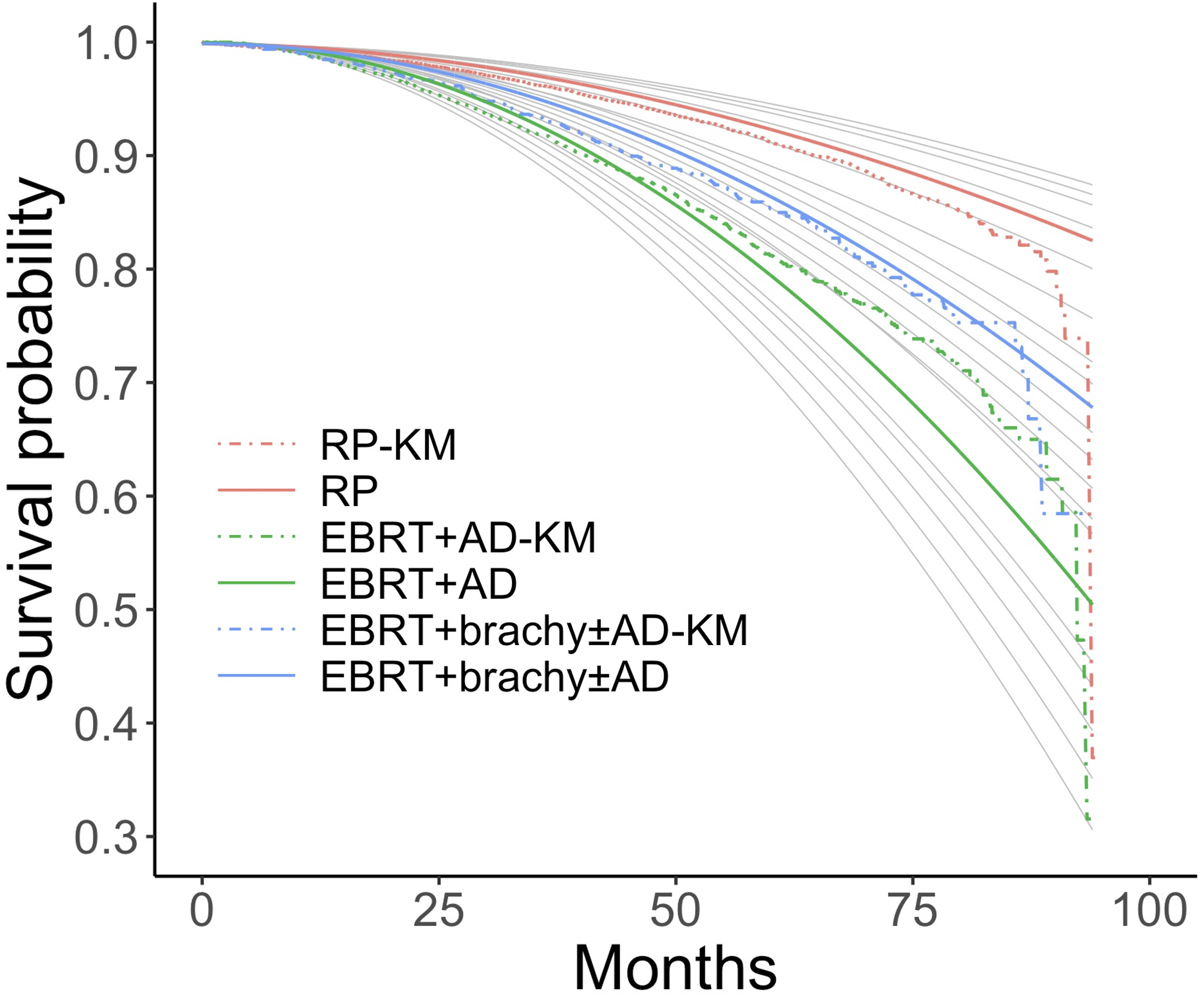}
\caption{The posterior mean of the counterfactual survival curves for each of three treatment groups in NCDB data. The solid curves are the average by treatment group of the individual-specific survival curves estimated following equation (7). The dashed survival curves are the Kaplan-Meier estimates for each treatment group. Solid gray curves are estimates of individual-specific survival curves for 15 randomly selected patients from three treatment groups.}
\label{fig:cf-surv}
\end{figure}

\begin{figure}[ht]
\centering 
\includegraphics[width = 1\textwidth]{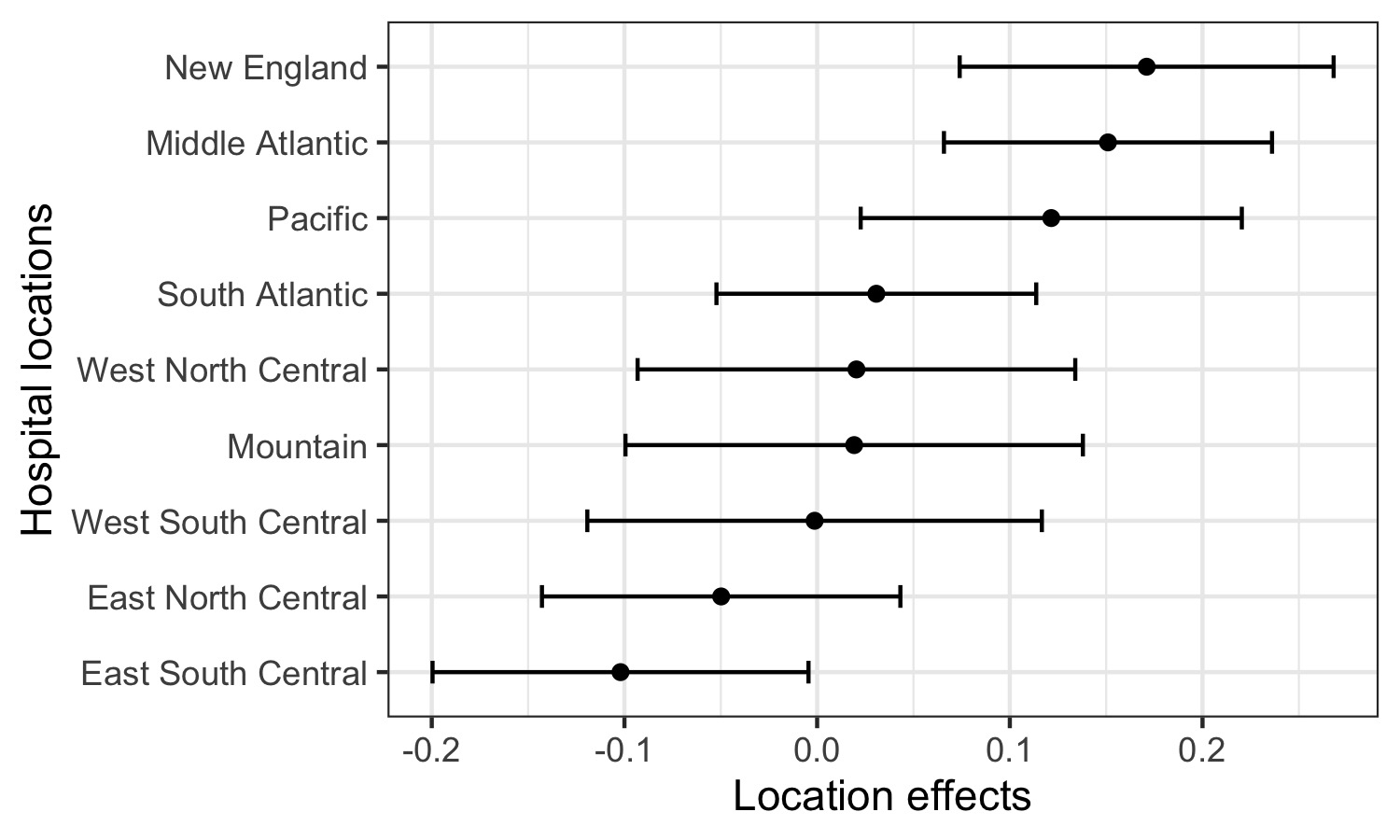}
\caption{The location effect in terms of the log survival time in months represented by the posterior mean and credible intervals of the random intercept $b_k$, $k=1, \ldots, 9$. }
\label{fig:ncdb_bk}
\end{figure}

Under the assumption of no unmeasured confounding, the treatment effect estimates, shown in Table \ref{tab:CATE-SA}, suggest that RP is the most beneficial treatment, which would on average lead to a ratio of 1.5 (1.3, 1.7) in the expected survival time compared to EBRT+AD and a ratio of 1.2 (1.1, 1.4)  compared to EBRT+brachy$\pm$AD. Between the two radiotherapy approaches, EBRT +AD leads to a shorter expected survival time that is 0.9 (0.7, 1) times the expected survival time for EBRT+brachy$\pm$AD.  Figure~\ref{fig:cf-surv}presents the posterior mean of the predicted counterfactual survival curves for each of three treatment groups. After correcting for confounding and accounting for the variability in location effects, our riAFT-BART estimators suggest increased treatment benefit associated with RP and reduced survival for EBRT+AD compared to the unadjusted Kaplan-Meier estimators. A perusal of the posterior distribution of the random intercepts $b_k$'s displayed in Figure~\ref{fig:ncdb_bk} suggests that there was substantial variability in the location effect. Hospitals in New England had significantly better outcomes (longer expected survival times) than hospitals in East Central area. 

It is possible that some important confounders were not collected in the NCDB data.
For example, patient functional status has been shown to be strongly associated with both treatment choices and survival among men with prostate cancer. \citep{jacobs2016association, stommel2002depression} The performance status measured by Eastern Clinical Oncology Group (ECOG) score may also be a confounder as it both affects the likelihood of clinicians choosing AD \citep{varenhorst2016predictors} and predicts survival for prostate cancer. \citep{lehtonen2020both} Futhermore, magnetic resonance imaging findings and the number of positive biopsy cores are both likely confounders as they are related to patient selection for RP or brachytherapy and indicate the degree of agressiveness of high-risk prostate cancer. \citep{chen2018challenges,nag1999american}

To evaluate the sensitivity of the treatment effect estimates to these unmeasured confounders, we first leverage the subject-area literature to specify the confounding functions. For the sign, we assume that  the unmeasured factors guiding clinicians to prescribe RP lead them systematically to prescribe it to relatively healthier patients. This is because magnetic resonance imaging findings supportive of resectability were used for patient selection for RP, \citep{chen2018challenges} and RP was recommended to those with lower number of positive biopsy cores, better functional score and better performance status. \citep{jacobs2016association, lee2000optimizing} Between the two radiotherapy based treatment approaches,  on the one hand, unhealthier patients with lower functional scores or ECOG scores were not recommended to use AD in treatment as they would not tolerate strong side effects induced by AD. \citep{varenhorst2016predictors, clay2007physical} On the other hand, brachytheray has been recommended as a boost to EBRT only to relatively unhealthier patients with multiple positive biopsy cores.\citep{nag1999american} Based on these pieces of evidence, we assume clinicians may have a preference to recommend EBRT+AD over EBRT+brachy$\pm$AD to healthier patients, but other directions of unmeasured confounding may also be plausible.

 We next postulate the bounds of confounding functions based on published scientific work. Lehtonen et al. \citep{lehtonen2020both} shows that ECOG score has a large effect (Cohen’s $d > 1$) and number of positive biopsy cores has a small effect (Cohen’s $d < 0.2$) on prostate cancer patient survival. Loosely translating Cohen’s $d$ to the proportion of the total variation in the outcome explained by a given covariate, \citep{lakens2013calculating} we assume the unmeasured confounders approximately account for 25\% of the total variance in overall survival. Following Chan et al., \citep{chan2018study} we fitted an AFT log-normal model to the NCDB data with fixed effects for locations, and computed the $R^2$ to be 60\%.  This suggests that the unmeasured confounders will explain no more than 40\% of the variation in the outcomes. Based on these grounds, we assume that the unmeasured confounding would account for  $\omega=0.75$ units of the remaining standard deviation unexplained by measured variables. The remaining standard deviation in the outcome unexplained by  measured covariates was estimated to be $\hat{\sigma}$ = 0.90 (months) via our riAFT-BART model. We hence set the bounds of the confounding functions to be $\pm \omega \hat{\sigma} = \pm .675$ (months). 

 Table~\ref{tab:CATE-SA} displays the sensitivity analysis results in comparison to treatment effect estimates under the weak unconfoundedness assumption.  We assume relatively healthier patients were assigned to RP:  $c(1,2\cond \bm{x},v) \sim \mathcal{U} (0, .675), \; c(2,1\cond \bm{x},v) \sim \mathcal{U}(-.675,0)$, and $c(1,3\cond \bm{x},v) \sim \mathcal{U} (0, .675), \; c(3,1\cond \bm{x},v) \sim \mathcal{U}(-.675,0)$. Between EBRT+AD and EBRT+brachy$\pm$AD, all four different forms of confounding functions are deemed plausible (see Web Table 1 for interpretations). 
 Results show that the significant treatment benefit associated with RP over EBRT+AD is robust to different magnitudes and directions of unmeasured confounding. However, the significant gain in the expected survival time  offered by RP in the causal analysis assuming weak unconfoundedness is negated in the presence of unmeasured confounding. Turning to the comparative effect between EBRT+AD and  EBRT+brachy$\pm$AD, under the assumption of no unmeasured confounding,  EBRT+brachy$\pm$AD had a survival benefit bordering on being statistically significant over EBRT +AD. Assuming unmeasured factors guiding clinicians to prescribe EBRT+AD lead them systematically to prescribe it to relatively healthier patients ``tips'' the result over to significant benefit for EBRT+brachy$\pm$AD. 
 
\begin{table}[!ht]
\centering
\caption{Estimation of causal effects of three treatment approaches on patient survival for high-risk localized prostate cancer, based on differences in log survival months, using the NCDB database. Three treatment options are $A = 1$: RP, $A = 2$: EBRT+AD and $A = 3$: EBRT+brachy$\pm$AD. Row 2--5 present confounding function adjusted effect estimates using our sensitivity analysis Algorithm~\ref{alg:SA}.  Interval estimates are based on pooled posterior samples across model fits arising from $30 \times 30$ data sets.   We assume relatively healthier patients were assigned to RP, $c(1,2\cond \bm{x},v)>0$, $c(2,1\cond \bm{x},v)<0$; $c(1,3\cond \bm{x},v)>0$, $c(3,1\cond \bm{x},v)<0$, and four unmeasured confounding directions between  $A = 2$: EBRT+AD and $A = 3$: EBRT+brachy$\pm$AD. We drew the values of confounding functions from the uniform distributions bounded by $\pm .675$ (months).}
\label{tab:CATE-SA}
\begin{tabular}{ccccc} 
\toprule
 & &$CATE_{1,2}$ & $CATE_{1,3}$ & $CATE_{2,3}$ \\
 \midrule
  &Assuming weak unconfoundedness &  $.38 (.24,.52)$ & $.22 (.07,.37)$ & $-.16 (-.32,.00)$\\
  \midrule 
  \multirow{4}{*}{Sensitivity analysis} &$c(2,3\cond \bm{x},v) >0, \; c(3,2\cond \bm{x},v) <0 $  & $.27 (.09,.45)$ & $.12 (-.07,.31)$ & $-.28(-.48,-.08)$\\
  & $c(2,3\cond \bm{x},v) <0, \; c(3,2\cond \bm{x},v) >0$  & $.31 (.13,.49)$ & $.13(-.06,.32)$ & $-.05 (-.25, .15)$\\
 &  $c(2,3\cond \bm{x},v) >0,\; c(3,2\cond \bm{x},v) >0$  & $.27 (.09,.45)$ & $.10 (-.09,.29)$ & $-.13 (-.33,.07)$\\
 &  $c(2,3\cond \bm{x},v) <0, \; c(3,2\cond \bm{x},v) <0$  & $.28 (.10,.46)$ & $.11 (-.08, .30)$ & $-.19 (-.39,.01)$\\
\bottomrule
\end{tabular}
\end{table}

\section{Summary and discussion} \label{sec:disc}
 The increased availability of large-scale healthcare databases allows researchers to conduct comparative effectiveness analysis of modern treatment approaches for high-risk cancer patients. Recent efforts have been made to investigate the effects of surgical and radiotherapy based treatments on patient survival using the national cancer databases. However, the multilevel data structure presented in these databases and its implications for causal analyses require special statistical consideration but have not been well studied. In addition, there is the lack of tools for evaluating the sensitivity of treatment effect estimates to the presence of unmeasured confounding in the context of multiple treatments and multilevel survival data. 
 
Motivated by these research gaps, our work makes two primary contributions to the causal inference literature. First,  we develop a flexible causal modeling tool and MCMC algorithm for causal inferences about  effects of multiple treatments on patient survival while respecting the multilevel data structure. The proposed riAFT-BART model flexibly captures the relationship between survival times and individual-level covariates and cluster-specific main effects, while providing proper representations of uncertainty intervals via the posterior based on a probability model. Second, leveraging the flexible riAFT-BART model, we develop an interpretable sensitivity analysis algorithm to address the causal assumption of no unmeasured confounding. In our sensitivity analysis approach, we present the confounding function on the basis of expected survival time, which can be easily interpreted, and propose methods to adjust the estimation of causal effects by effectively removing the bias due to posited levels of unmeasured confounding. 

Applying our proposed methods to NCDB data on older high-risk localized prostate cancer patients, we confirmed the survival benefit of RP relative to EBRT+AD. Inferences about other two pairwise treatment effects were inconclusive because they were impacted by the potential unmeasured confounding. Our causal analysis also demonstrates that there is substantial variability in the effects of hospital locations, which reinforces the importance of examining the cluster-level variation when estimating treatment effect in the general population using data with hierarchical structure. 

There are several important avenues for future research. First, our riAFT-BART model can be extended to include the random slopes and accommodate the cluster-level covariates. Second, developing a sensitivity analysis for cluster-level unmeasured confounding could be a worthwhile and important contribution. Third, although our simulation results suggest that our methods are robust to the normality assumption for the random intercepts and residuals, it may be worthwhile to develop nonparametric priors to further improve the modeling flexibility. Finally, to address the causal assumption of positivity, we can extend the work by Hill and Su \citep{hill2013assessing} and Hu et al. \citep{hu2020estimation} to develop a strategy to identify a common support region for inferential units. 

\section*{ACKNOWLEDGEMENTS}
This work was supported in part by the National Cancer Institute under grant NIH NCI R21CA245855, and by award ME\_2017C3\_9041 from the Patient-Centered Outcomes Research Institute (PCORI). The content is solely the responsibility of the authors and does not necessarily represent the official views of the National Institutes of Health or PCORI. %The authors thank the Associate Editor and two anonymous referees, whose suggestions have substantially improved the exposition of this work. 

\section*{DATA AVAILABILITY STATEMENT}
The simulation codes that generate the data supporting the findings of the simulation study are openly available with the $\R$ package $\code{riAFTBART}$. 
The NCDB data used in the case study is publicly available upon approval of the NCDB Participant User File application. 
\bibliography{reference}

\begin{thebibliography}{10}
\providecommand \doibase [0]{http://dx.doi.org/}%

\bibitem{ennis2018brachytherapy}
Ennis RD, Hu L, Ryemon SN, Lin J, Mazumdar M. Brachytherapy-based radiotherapy
  and radical prostatectomy are associated with similar survival in high-risk
  localized prostate cancer. {\it Journal of Clinical Oncology} 2018\string;
  36(12)\string: 1192--1198.

\bibitem{chen2018challenges}
Chen RC. Challenges of interpreting registry data in prostate cancer:
  interpreting retrospective results along with or in absence of clinical trial
  data. {\it Journal of Clinical Oncology} 2018\string; 36(12)\string:
  1181--1183.

\bibitem{hu2021estimating}
Hu L, Ji J, Li F. Estimating heterogeneous survival treatment effect in
  observational data using machine learning. {\it Statistics in Medicine}
  2021\string; 40(21)\string: 4691--4713.

\bibitem{zeng2022propensity}
Zeng S, Li F, Hu L, Li F. {Propensity Score Weighting Analysis of Survival
  Outcomes Using Pseudo-observations}. {\it Statistica Sinica} 2022.
\newblock In press.

\bibitem{feng2012generalized}
Feng P, Zhou XH, Zou QM, Fan MY, Li XS. Generalized propensity score for
  estimating the average treatment effect of multiple treatments. {\it
  Statistics in Medicine} 2012\string; 31(7)\string: 681--697.

\bibitem{mccaffrey2013tutorial}
McCaffrey DF, Griffin BA, Almirall D, Slaughter ME, Ramchand R, Burgette LF. A
  tutorial on propensity score estimation for multiple treatments using
  generalized boosted models. {\it Statistics in Medicine} 2013\string;
  32(19)\string: 3388--3414.

\bibitem{linden2016estimating}
Linden A, Uysal SD, Ryan A, Adams JL. Estimating causal effects for multivalued
  treatments: a comparison of approaches. {\it Statistics in Medicine}
  2016\string; 35(4)\string: 534--552.

\bibitem{hu2020estimation}
Hu L, Gu C, Lopez M, Ji J, Wisnivesky J. Estimation of causal effects of
  multiple treatments in observational studies with a binary outcome. {\it
  Statistical Methods in Medical Research} 2020\string; 29(11)\string:
  3218--3234.

\bibitem{hu2021estimation}
Hu L, Gu C. Estimation of causal effects of multiple treatments in healthcare
  database studies with rare outcomes. {\it Health Services and Outcomes
  Research Methodology} 2021\string; 21(3)\string: 287--308.

\bibitem{yu2021comparison}
Yu Y, Zhang M, Shi X, Caram ME, Little RJ, Mukherjee B. A comparison of
  parametric propensity score-based methods for causal inference with multiple
  treatments and a binary outcome. {\it Statistics in Medicine} 2021\string;
  40(7)\string: 1653--1677.

\bibitem{von2007strengthening}
Von~Elm E, Altman DG, Egger M, Pocock SJ, G{\o}tzsche PC, Vandenbroucke JP. The
  Strengthening the Reporting of Observational Studies in Epidemiology (STROBE)
  statement: guidelines for reporting observational studies. {\it Bulletin of
  the World Health Organization} 2007\string; 85(11)\string: 867--872.

\bibitem{hu2022flexible}
Hu L, Zou J, Gu C, Ji J, Lopez M, Kale M. A flexible sensitivity analysis
  approach for unmeasured confounding with multiple treatments and a binary
  outcome with application to SEER-Medicare lung cancer data. {\it The Annals
  of Applied Statistics} 2022.
\newblock In press.

\bibitem{chipman2010bart}
Chipman HA, George EI, McCulloch RE. {BART: Bayesian additive regression
  trees}. {\it The Annals of Applied Statistics} 2010\string; 4(1)\string:
  266--298.

\bibitem{hill2011bayesian}
Hill JL. Bayesian nonparametric modeling for causal inference. {\it Journal of
  Computational and Graphical Statistics} 2011\string; 20(1)\string: 217--240.

\bibitem{hu2021variable}
Hu L, Lin J, Ji J. {Variable selection with missing data in both covariates and
  outcomes: Imputation and machine learning}. {\it Statistical Methods in
  Medical Research} 2021\string; 30(12)\string: 2651--2671.

\bibitem{hu2021est}
Hu L, Lin JY, Sigel K, Kale M. Estimating heterogeneous survival treatment
  effects of lung cancer screening approaches: A causal machine learning
  analysis. {\it Annals of Epidemiology} 2021\string; 62\string: 36--42.

\bibitem{chen2001causal}
Chen PY, Tsiatis AA. Causal inference on the difference of the restricted mean
  lifetime between two groups. {\it Biometrics} 2001\string; 57(4)\string:
  1030--1038.

\bibitem{arpino2016propensity}
Arpino B, Cannas M. Propensity score matching with clustered data. An
  application to the estimation of the impact of caesarean section on the Apgar
  score. {\it Statistics in Medicine} 2016\string; 35(12)\string: 2074--2091.

\bibitem{hernan2006estimating}
Hern{\'a}n MA, Robins JM. Estimating causal effects from epidemiological data.
  {\it Journal of Epidemiology \& Community Health} 2006\string; 60(7)\string:
  578--586.

\bibitem{hernan2020causal}
Hern{\'a}n MA, Robins JM. {\it Causal Inference: What If}.
\newblock Boca Raton: Chapman \& Hall/CRC .
\newblock 2020.

\bibitem{tan2019bayesian}
Tan YV, Roy J. Bayesian additive regression trees and the General BART model.
  {\it Statistics in Medicine} 2019\string; 38(25)\string: 5048--5069.

\bibitem{gelman2008using}
Gelman A, Van~Dyk DA, Huang Z, Boscardin JW. Using redundant parameterizations
  to fit hierarchical models. {\it Journal of Computational and Graphical
  Statistics} 2008\string; 17(1)\string: 95--122.

\bibitem{henderson2020individualized}
Henderson NC, Louis TA, Rosner GL, Varadhan R. Individualized treatment effects
  with censored data via fully nonparametric {Bayesian} accelerated failure
  time models. {\it Biostatistics} 2020\string; 21(1)\string: 50--68.

\bibitem{roy2017bayesian}
Roy J, Lum KJ, Daniels MJ. {A Bayesian nonparametric approach to marginal
  structural models for point treatments and a continuous or survival outcome}.
  {\it Biostatistics} 2017\string; 18(1)\string: 32--47.

\bibitem{rosenbaum2002covariance}
Rosenbaum PR. Covariance adjustment in randomized experiments and observational
  studies. {\it Statistical Science} 2002\string; 17(3)\string: 286--327.

\bibitem{kasza2017assessing}
Kasza J, Wolfe R, Schuster T. Assessing the impact of unmeasured confounding
  for binary outcomes using confounding functions. {\it International Journal
  of Epidemiology} 2017\string; 46(4)\string: 1303--1311.

\bibitem{robins1999association}
Robins JM. Association, causation, and marginal structural models. {\it
  Synthese} 1999\string: 151--179.

\bibitem{vanderweele2017sensitivity}
VanderWeele TJ, Ding P. {Sensitivity analysis in observational research:
  introducing the E-value}. {\it Annals of Internal Medicine} 2017\string;
  167(4)\string: 268--274.

\bibitem{arpino2011specification}
Arpino B, Mealli F. The specification of the propensity score in multilevel
  observational studies. {\it Computational Statistics \& Data Analysis}
  2011\string; 55(4)\string: 1770--1780.

\bibitem{li2013propensity}
Li F, Zaslavsky AM, Landrum MB. Propensity score weighting with multilevel
  data. {\it Statistics in Medicine} 2013\string; 32(19)\string: 3373--3387.

\bibitem{fuentes2021causal}
Fuentes A, L{\"u}dtke O, Robitzsch A. {Causal Inference with Multilevel Data: A
  Comparison of Different Propensity Score Weighting Approaches}. {\it
  Multivariate Behavioral Research} 2021.
\newblock In press.

\bibitem{brumback2004sensitivity}
Brumback BA, Hern{\'a}n MA, Haneuse SJ, Robins JM. Sensitivity analyses for
  unmeasured confounding assuming a marginal structural model for repeated
  measures. {\it Statistics in Medicine} 2004\string; 23(5)\string: 749--767.

\bibitem{rubin2003nested}
Rubin DB. {Nested multiple imputation of NMES via partially incompatible MCMC}.
  {\it Statistica Neerlandica} 2003\string; 57(1)\string: 3--18.

\bibitem{zhou2010note}
Zhou X, Reiter JP. {A note on Bayesian inference after multiple imputation}.
  {\it The American Statistician} 2010\string; 64(2)\string: 159--163.

\bibitem{hogan2014bayesian}
Hogan JW, Daniels MJ, Hu L. A Bayesian perspective on assessing sensitivity to
  assumptions about unobserved data. In:  Molenberghs G, Fitzmaurice G, Kenward
  MG, Tsiatis A, Verbeke G. \kern-2pt, eds. {\it Handbook of Missing Data
  Methodology}Boca Raton, FL: CRC Press.  2014 (pp. 405--434).

\bibitem{hastie1990generalized}
Hastie TJ, Tibshirani RJ. {\it Generalized additive models}.
\newblock Boca Raton, FL: Chapman \&Hall .
\newblock 1990.

\bibitem{royston2013restricted}
Royston P, Parmar MK. {Restricted mean survival time: an alternative to the
  hazard ratio for the design and analysis of randomized trials with a
  time-to-event outcome}. {\it BMC Medical Research Methodology} 2013\string;
  13(1)\string: 1--15.

\bibitem{cai2011additive}
Cai J, Zeng D. Additive mixed effect model for clustered failure time data.
  {\it Biometrics} 2011\string; 67(4)\string: 1340--1351.

\bibitem{van2007super}
Laan V.~dMJ, Polley EC, Hubbard AE. Super learner. {\it Statistical
  Applications in Genetics and Molecular Biology} 2007\string; 6(1).

\bibitem{jacobs2016association}
Jacobs BL, Lopa SH, Yabes JG, Nelson JB, Barnato AE, Degenholtz HB. Association
  of functional status and treatment choice among older men with prostate
  cancer in the Medicare Advantage population. {\it Cancer} 2016\string;
  122(20)\string: 3199--3206.

\bibitem{stommel2002depression}
Stommel M, Given BA, Given CW. Depression and functional status as predictors
  of death among cancer patients. {\it Cancer} 2002\string; 94(10)\string:
  2719--2727.

\bibitem{varenhorst2016predictors}
Varenhorst E, Klaff R, Berglund A, Hedlund PO, Sandblom G, 5 SPCGSTN.
  Predictors of early androgen deprivation treatment failure in prostate cancer
  with bone metastases. {\it Cancer Medicine} 2016\string; 5(3)\string:
  407--414.

\bibitem{lehtonen2020both}
Lehtonen M, Heiskanen L, Reinikainen P, Kellokumpu-Lehtinen PL. Both
  comorbidity and worse performance status are associated with poorer overall
  survival after external beam radiotherapy for prostate cancer. {\it BMC
  Cancer} 2020\string; 20(1)\string: 1--8.

\bibitem{nag1999american}
Nag S, Beyer D, Friedland J, Grimm P, Nath R. American Brachytherapy Society
  (ABS) recommendations for transperineal permanent brachytherapy of prostate
  cancer. {\it International Journal of Radiation Oncology, Biology, Physics}
  1999\string; 44(4)\string: 789--799.

\bibitem{lee2000optimizing}
Lee A, Schultz D, Renshaw A, Richie J, D’Amico A. Optimizing patient
  selection for prostate monotherapy. {\it International Journal of Radiation
  Oncology, Biology, Physics} 2000\string; 3(48)\string: 306--307.

\bibitem{clay2007physical}
Clay CA, Perera S, Wagner JM, Miller ME, Nelson JB, Greenspan SL. Physical
  function in men with prostate cancer on androgen deprivation therapy. {\it
  Physical Therapy} 2007\string; 87(10)\string: 1325--1333.

\bibitem{lakens2013calculating}
Lakens D. Calculating and reporting effect sizes to facilitate cumulative
  science: a practical primer for t-tests and ANOVAs. {\it Frontiers in
  psychology} 2013\string; 4\string: 863.

\bibitem{chan2018study}
Chan PH, Xu R, Chambers CD. {A study of $R^2$ measure under the accelerated
  failure time models}. {\it Communications in Statistics-Simulation and
  Computation} 2018\string; 47(2)\string: 380--391.

\bibitem{hill2013assessing}
Hill J, Su YS. {Assessing lack of common support in causal inference using
  Bayesian nonparametrics: Implications for evaluating the effect of
  breastfeeding on children's cognitive outcomes}. {\it The Annals of Applied
  Statistics} 2013\string; 7(3)\string: 1386--1420.

\end{thebibliography}
\end{document}

\title{\Large Web-based Supplementary Materials for ``A flexible approach for causal inference with multiple treatments and clustered survival outcomes" by Hu et al.}

\author[1]{\emph{email}: liangyuan.hu@rutgers.edu}
\date{}
\maketitle

\beginsupplement

\section{Posterior distributions of $\mu_{lh}$, $\sigma^2$, $b_k$, $\alpha_k$ and $\tau^2$ in riAFT-BART}

\textbf{1) For the posterior of $\sigma^{2},$} since we have $\sigma^{2}\sim IG\left(\frac{\nu}{2},\frac{\nu\lambda}{2}\right),$
we obtain
\begin{align*}
 & \hspace{12pt}P\lp \sigma^2 \cond  y^{cent,c}_{ik}, \bm{X}_{ik}, A_{ik}, V_k, b_k, \tau^2, \alpha_k, \{\mathcal{W}_H, \mathcal{M}_H\}\rp \\
 & \propto P \left( y^{cent,c}_{ik} \cond \bm{X}_{ik}, A_{ik}, V_k,b_k, \{\mathcal{W}_H, \mathcal{M}_H\}\right) P \left(\sigma^2 \right) \\
 & \propto\left\{ \prod_{k=1}^{K}\prod_{i=1}^{n_{k}}\left(\sigma^{2}\right)^{-\frac{1}{2}}\exp\left[-\frac{\lp y^{cent,c}_{ik} - \hat{f}(\bm{X}_{ik}, A_{ik})-b_k\rp^2}{2\sigma^{2}}\right]\right\} \left(\sigma^{2}\right)^{-\left(\frac{\nu}{2}+1\right)}\exp\left[-\frac{\nu\lambda}{2\sigma^{2}}\right]\\
 & \propto\left(\sigma^{2}\right)^{-\left(\frac{N+\nu}{2}+1\right)}\exp\left[-\frac{\sum_{k=1}^{K}\sum_{i=1}^{n_{k}}\lp y^{cent,c}_{ik} - \hat{f}(\bm{X}_{ik}, A_{ik})-b_k\rp^2+\nu\lambda}{2\sigma^{2}}\right]
\end{align*}
\\
\begin{align*}
&P\left(\sigma^2  \cond y^{cent,c}_{ik}, \bm{X}_{ik}, A_{ik}, V_k, \tau^2, \alpha_k, b_k, \{\mathcal{W}_H, \mathcal{M}_H\}\right) \\
&\sim IG\left(\frac{N+\nu}{2},\frac{\sum_{k=1}^{K}\sum_{i=1}^{n_{k}}\lp y^{cent,c}_{ik} - \hat{f}(\bm{X}_{ik}, A_{ik})-b_k\rp^2+\nu\lambda}{2}\right)
\end{align*}
\\

\textbf{2) For the posterior of the random intercept $b_k$}  
 \begin{align*}
    & \hspace{12pt}P \left(b_k \cond y^{cent,c}_{ik}, \bm{X}_{ik}, A_{ik}, V_k, \tau^2, \alpha_k, \sigma^2, \{\mathcal{W}_H, \mathcal{M}_H\} \right) \\
    & \propto P \left( y^{cent,c}_{ik} \cond \bm{X}_{ik}, A_{ik}, V_k, \{\mathcal{W}_H, \mathcal{M}_H\}, \sigma^2 , b_k\right) P \left(b_k \cond \tau^2, \alpha_k \right)\\
    & \propto \lsq \prod_{i=1}^{n_k} \exp \lbc - \dfrac{\lp y^{cent,c}_{ik} - \hat{f}(\bm{X}_{ik}, A_{ik})-b_k\rp^2}{2\sigma^2} \rbc \rsq \exp \lbc  -\dfrac{b_k^2}{\alpha_k\tau^2}\rbc\\
    & \propto\exp\lbc-\frac{\sum_{i=1}^{n_{k}}\lp y^{cent,c}_{ik} - \hat{f}(\bm{X}_{ik}, A_{ik})-b_k\rp^2}{2\sigma^{2}}\rbc \exp\lbc -\frac{b_{k}^{2}}{2{\alpha_{k}}\tau^{2}}\rbc\\
    & \propto \exp\lbc-\frac{\left(n_{k}\tau^{2}{\alpha_{k}}+\sigma^{2}\right)b_{k}^{2}-2\tau^{2}b_{k}{\alpha_{k}}\sum_{i=1}^{n_{k}}\lp y^{cent,c}_{ik} - \hat{f}(\bm{X}_{ik}, A_{ik})\rp}{2\sigma^{2}\tau^{2}{\alpha_{k}}}\rbc\\
& \propto \exp \lbc - \dfrac{\lp b_k - \dfrac{\tau^2 \alpha_k \sum_{i=1}^{n_k}\lp y^{cent,c}_{ik}- \hat{f}(\bm{X}_{ik}, A_{ik})\rp}{n_k\tau^2\alpha_k+\sigma^2}\rp^2}{2\dfrac{\sigma^2 \tau^2 \alpha_k}{n_k \tau^2\alpha_k+\sigma^2}} \rbc.
 \end{align*}
 
 \[
P \left(b_k \cond y^{cent,c}_{ik}, \bm{X}_{ik}, A_{ik}, V_k, \tau^2, \alpha_k, \sigma^2, \{\mathcal{W}_H, \mathcal{M}_H\} \right) \sim N\left(\dfrac{\tau^2 \alpha_k \sum_{i=1}^{n_k}\lp y^{cent,c}_{ik}- \hat{f}(\bm{X}_{ik}, A_{ik})\rp}{n_k\tau^2\alpha_k+\sigma^2},\frac{\sigma^{2}\tau^{2}{\alpha_{k}}}{n_{k}\tau^{2}{\alpha_{k}}+\sigma^{2}}\right)
\]
 
\textbf{3) For the posterior of $\alpha_k$,  used for parameter expansion}
 \begin{align*}
    &\hspace{12pt}P\left(\alpha_k \cond y^{cent,c}_{ik}, \bm{X}_{ik}, A_{ik}, V_k, \tau^2, b_k, \sigma^2, \{\mathcal{W}_H, \mathcal{M}_H\} \right)\\
    &\propto \lbc\prod_{k=1}^K P\lp b_k \cond \tau^2, \alpha_k \rp  \rbc P(\alpha_k)\\
    &\propto \exp \lbc -\frac{\sum_{k=1}^K b_k^2}{2\alpha_k\tau^2} \rbc \lbc \alpha_k^{-2} \exp\lp -\frac{1}{\alpha_k}\rp \rbc\\
    &\propto \alpha_k^{-(1+1)}\exp \lbc -\frac{1}{\alpha_k} \lp 1+\frac{\sum_{k=1}^K b_k^2}{2\tau^2} \rp \rbc.
\end{align*}
\\
\[
P\left(\alpha_k \cond y^{cent,c}_{ik}, \bm{X}_{ik}, A_{ik}, V_k, \tau^2, b_k, \sigma^2, \{\mathcal{W}_H, \mathcal{M}_H\} \right)\sim IG\left(1,1+\frac{\sum_{k=1}^K b_k^2}{2\tau^2}\right)
\]
\\

\textbf{4) For the posterior of $\tau^2$}
\begin{align*}
    & \hspace{12pt}P\left( \tau^2 \cond  y^{cent,c}_{ik}, \bm{X}_{ik}, A_{ik}, V_k,  b_k, \alpha_k, \sigma^2, \{\mathcal{W}_H, \mathcal{M}_H\} \right)\\
    &\propto P\left( y^{cent,c}_{ik} \cond \bm{X}_{ik}, A_{ik}, V_k, \{\mathcal{W}_H, \mathcal{M}_H\}, \sigma^2, b_k, \alpha_k \right) P(\tau^2)\\
    &\propto \lbc \prod_{k=1}^K P\lp b_k \cond \tau^2, \alpha_k \rp \rbc P(\tau^2)\\
    & \propto (\tau^2)^{-K/2} \exp \lbc -\frac{\sum_{k=1}^K b_k^2}{2\alpha_k\tau^2} \rbc (\tau^2)^{-2} \exp\lp -\frac{1}{\tau^2} \rp\\
    &\propto (\tau^2)^{-K/2+1+1}\exp \lp -\frac{\sum_{k=1}^K b_k^2+2\alpha_k}{2\alpha_k\tau^2} \rp.
\end{align*}
\\
\[
P\left( \tau^2 \cond  y^{cent,c}_{ik}, \bm{X}_{ik}, A_{ik}, V_k,  b_k, \alpha_k, \sigma^2, \{\mathcal{W}_H, \mathcal{M}_H\} \right)\sim IG\left(\frac{K}{2}+1,\frac{\sum_{k=1}^{K}b_{k}^{2}+2{\alpha_{k}}}{2{\alpha_{k}}}\right)
\]
\\

\textbf{5) For the posterior of $\mu_{lh}$}, we first draw the Gibbs
sample of $b_{k},\alpha_{i},\tau^{2},\sigma^{2}$ seperately from
their respective posterior distribution. Then using the updated $b_{k},$
we obtain $\tilde{Y}_{ik}^{c,c}=Y_{ik}^{c,c}-b_{k}.$ Now $\tilde{Y}_{ik}^{c,c}\mid\boldsymbol{X}_{ik}$
can be viewed as a BART model. We have $\tilde{R}_{lh}\mid g\left(\mathbf{X}_{ik},T_{j},\mathbf{M}_{j}\right),\sigma\sim N(\mu_{lh},\sigma^{2}),\mu_{lh}\mid \{\mathcal{W}_H\}\sim N(\mu_{\mu},\sigma_{\mu}^{2})$
where $\tilde{R}_{lh}^{c,c}=Y_{ik}^{c,c}-b_{k}-\sum_{w\neq h}g\left(\mathbf{X}_{ik},\mathcal{W}_w, \mathcal{M}_w\right)$
\begin{align*}
& \hspace{12pt}P \lp \mu_{lh} \cond y^{cent,c}_{ik}, \bm{X}_{ik}, A_{ik}, V_k, b_k, \tau^2, \alpha_k, \sigma^2, \{\mathcal{W}_H\} \rp \\
& \propto  P(\mu_{lh}\mid \{\mathcal{W}_H\},\sigma^2,\tilde{R}_{lh})\\
& \propto P(\tilde{R}_{lh}\mid \{\mathcal{W}_H\},\mu_{lh},\sigma)P(\mu_{lh}\mid \{\mathcal{W}_H\})\\
& \propto\exp\lbc-\frac{\sum_{i}(\tilde{r}_{lh}-\mu_{lh})^{2}}{2\sigma^{2}}\rbc\exp\lbc-\frac{(\mu_{lh}-\mu_{\mu})^{2}}{2\sigma_{\mu}^{2}}\rbc\\
& \propto\exp\lbc-\frac{\sigma_{\mu}^{2}\sum_{i}(\tilde{r}_{lh}^{2}+\mu_{lh}-2\mu_{lh}\tilde{r}_{lh})+\sigma^{2}(\mu_{lh}^{2}+\mu_{\mu}^{2}-2\mu_{\mu}\mu_{lh})}{2\sigma^{2}\sigma_{\mu}^{2}}\rbc\\
& \propto\exp\lbc-\frac{\sigma_{\mu}^{2}n_{i}\mu_{lh}-2\sigma_{\mu}^{2}\mu_{lh}\sum_{i}\tilde{r}_{lh}+\sigma^{2}\mu_{lh}^{2}-2\sigma^{2}\mu_{\mu}\mu_{lh})}{2\sigma^{2}\sigma_{\mu}^{2}}\rbc\\
& \propto\exp\lbc-\frac{(n_{i}\sigma_{\mu}^{2}+\sigma^{2})\mu_{lh}^{2}-2(\sigma_{\mu}^{2}\sum_{i}\tilde{r}_{lh}+\sigma^{2}\mu_{\mu})\mu_{lh}}{2\sigma^{2}\sigma_{\mu}^{2}}\rbc\\
& \propto\exp\lbc-\frac{(\mu_{lh}-\frac{\sigma_{\mu}^{2}\sum_{i}\tilde{r}_{lh}+\sigma^{2}\mu_{\mu}}{n_{i}\sigma_{\mu}^{2}+\sigma^{2}})^{2}}{\frac{2\sigma^{2}\sigma_{\mu}^{2}}{n_{i}\sigma_{\mu}^{2}+\sigma^{2}}}\rbc
\end{align*}
\[
P \lp \mu_{lh} \cond y^{cent,c}_{ik}, \bm{X}_{ik}, A_{ik}, V_k, b_k, \tau^2, \alpha_k, \sigma^2, \{\mathcal{W}_H\} \rp\sim N\left(\frac{\sigma_{\mu}^{2}\sum_{i}\tilde{r}_{lh}+\sigma^{2}\mu_{\mu}}{n_{i}\sigma_{\mu}^{2}+\sigma^{2}},\frac{\sigma^{2}\sigma_{\mu}^{2}}{n_{i}\sigma_{\mu}^{2}+\sigma^{2}}\right)
\]

\section{Technical details for sensitivity analysis in Section 3.2}
The derivation of the bias formula $\text{Bias}(a_j,a_{j'} \cond \bm{x},v)$ in  equation (5) follows the proof of Theorem 2.1 in \cite{hu2022flexible}.  

Under the weak unconfoundedness assumption (A2), ignoring individual-level unmeasured confounding will lead to the following bias in the estimate of the causal effect $CATE_{a_j,a_{j'}}$, 
\begin{eqnarray*}
\text{Bias}(a_j,a_{j'} \cond \bm{x},v) &=& E[T|A=a_j, \bm{X}=\bm{x},V=v] -E[T|A=a_{j'}, \bm{X}=\bm{x},V=v]\\
&&-E\lsq T(a_j)-T(a_{j'})\cond \bm{X}=\bm{x},V=v \rsq. 
\end{eqnarray*}
To simplify notation, we will use $E\lsq \cdot \cond a, \bm{x}, v\rsq $ to denote $E\lsq \cdot \cond A=a, \bm{X}= \bm{x}, V=v\rsq $. 
Applying the law of total expectation to $E\lsq T(a_j \cond \bm{X}=\bm{x}, V=v)\rsq$, we have 
\begin{eqnarray*}
E\lsq T(a_j) \cond \bm{X}=\bm{x}, V=v \rsq &=& \sum_{l=0}^J p_{l}E\lsq T(a_j)\cond a_l, \bm{X}=\bm{x}, V=v \rsq. 
\end{eqnarray*}
Similary, 
\begin{eqnarray*}
E\lsq T(a_{j'}) \cond \bm{X}=\bm{x}, V=v \rsq &=& \sum_{l=0}^J p_{l}E\lsq T(a_{j
})\cond \bm{X}=\bm{x}, V=v \rsq. 
\end{eqnarray*}
Hence, 
\begin{eqnarray} \nonumber
E \lsq T(a_j) - Y(a_{j'}) \cond \bm{x}, v  \rsq & = &p_0 E \lsq T(a_j)-T(a_{j'}) \cond a_0, \bm{x}, v  \rsq + \ldots + p_J E \lsq T(a_j)-T(a_{j'}) \cond a_J, \bm{x}, v  \rsq  \\ \label{eq:twoEs}
     &&+ p_j E \lsq T(a_j)-T(a_{j'}) \cond a_j, \bm{x}, v  \rsq +  p_{j'} E \lsq T(a_j)-T(a_{j'}) \cond a_{j'}, \bm{x}, v  \rsq.  
\end{eqnarray}

Repeatedly using $E \lsq T(a_l) \cond a_l, \bm{x}, v \rsq = E \lsq T \cond a_l, \bm{x}, v \rsq,  \forall l \in \{1, \ldots, J\}$ to rewrite the last two items of the RHS of the equation~\eqref{eq:twoEs}, we have  
\begin{eqnarray}\nonumber
&&p_j E \lsq T(a_j)-T(a_{j'}) \cond a_j,\bm{x}, v \rsq +  p_{j'} E \lsq T(a_j)-T(a_{j'}) \cond a_{j'}, \bm{x}, v \rsq\\\nonumber
&=&p_j E \lsq T(a_j) \cond a_j, \bm{x}, v \rsq - p_j E \lsq T(a_{j'}) \cond a_j, \bm{x}, v \rsq + p_{j'} E \lsq T(a_j) \cond a_{j'}, \bm{x}, v \rsq - p_{j'} E \lsq T(a_{j'}) \cond a_{j'}, \bm{x}, v \rsq \\\nonumber
&=& p_j E \lsq T \cond a_j, \bm{x}, v \rsq - p_{j'} E \lsq T \cond a_{j'}, \bm{x}, v \rsq + p_{j'} E \lsq T (a_j)\cond a_{j'}, \bm{x}, v \rsq - p_j E \lsq T (a_{j'})\cond a_j, \bm{x}, v \rsq  \\\nonumber
&=& p_j \lbc E \lsq T \cond a_j, \bm{x}, v \rsq -  E \lsq  T \cond a_{j'}, \bm{x}, v \rsq \rbc + p_j  E \lsq  T \cond a_{j'}, \bm{x}, v \rsq - p_{j'} E \lsq T \cond a_{j'}, \bm{x}, v \rsq \\\nonumber
&& - p_j E \lsq T (a_{j'})\cond a_j, \bm{x}, v \rsq + p_{j'} E \lsq T (a_j)\cond a_{j'}, \bm{x}, v \rsq\\\nonumber
&=&  p_j \lbc E \lsq T \cond a_j, \bm{x}, v \rsq -  E \lsq  T \cond a_{j'}, \bm{x}, v \rsq \rbc + p_j \lbc  E \lsq T(a_{j'}) \cond a_{j'}, \bm{x}, v \rsq - E \lsq T(a_{j'}) \cond a_j, \bm{x}, v \rsq  \rbc \\\nonumber
&&+ p_{j'} \lbc  E \lsq  T(a_j) - T(a_{j'}) \cond a_{j'}, \bm{x}, v \rsq \rbc \\ 
&=&p_j \lbc E \lsq T \cond a_j, \bm{x}, v \rsq -  E \lsq  T \cond a_{j'}, \bm{x}, v \rsq \rbc +p_j c(a_{j'}, a_j, \bm{x}, v) +  p_{j'} \lbc  E \lsq  T(a_j) - T(a_{j'}) \cond a_{j'}, \bm{x}, v \rsq \rbc. 
\label{eq: eqpk}
\end{eqnarray}

Let $\tilde{p} = 1- p_j - p_{j'}$.  By rewriting $p_{j'} \lbc  E \lsq  T(a_j) - T(a_{j'}) \cond a_{j'}, \bm{x}, v\rsq \rbc$  in equation~\eqref{eq: eqpk}, we have 
\begin{eqnarray*}
&&p_{j'} \lbc  E \lsq  T(a_j) - T(a_{j'}) \cond a_{j'}, \bm{x}, v \rsq \rbc \\
&=& (1-p_j-\tilde{p}) \lbc E \lsq T \cond a_j , \bm{x}, v  \rsq - E \lsq T \cond a_{j'}, \bm{x}, v \rsq  + E \lsq  T(a_j) \cond a_{j'}, \bm{x}, v \rsq  - E \lsq  T(a_j) \cond a_j, \bm{x}, v \rsq \rbc\\
&=& (1-p_j)  \lbc E \lsq T \cond a_j , \bm{x}, v  \rsq - E \lsq T \cond a_{j'}, \bm{x}, v \rsq \rbc  - (1-p_j - \tilde{p}) c(a_j, a_{j'}, \bm{x}, v)  \\
&&- \tilde{p} \lbc E \lsq T \cond a_j, \bm{x}, v \rsq - E \lsq T \cond a_{j'}, \bm{x}, v \rsq  \rbc \\
&=& (1-p_j)  \lbc E \lsq T \cond a_j , \bm{x}, v  \rsq - E \lsq T \cond a_{j'}, \bm{x}, v \rsq \rbc - p_{j'} c(a_j, a_{j'}, \bm{x}, v) \\
&&- (1-p_j-p_{j'}) \lbc E \lsq T (a_j) \cond a_j, \bm{x}, v \rsq - E \lsq T(a_{j'}) \cond a_{j'}, \bm{x}, v \rsq  \rbc.
\end{eqnarray*}

Taken together, 
\begin{eqnarray*}
\text{Bias}(a_j,a_{j'} \cond \bm{x},v) &= & \sum_{m: a_m \in \mathscr{A} \setminus\{a_j, a_{j'}\} } -p_m E \lsq  T(a_j) - T(a_{j'}) \cond a_m, \bm{x}, v \rsq - p_j c(a_{j'}, a_j, \bm{x}, v) + p_{j'} c(a_j, a_{j'}, \bm{x}, v) \\
&& + \sum_{m: a_m \in \mathscr{A} \setminus\{a_j, a_{j'}\} } p_m\lbc  E\lsq T(a_j) \cond a_j, \bm{x}, v \rsq  - E\lsq T(a_{j'}) \cond a_{j'}, \bm{x}, v \rsq \rbc\\
&=& -p_j c(a_{j'}, a_j, \bm{x}, v)+ p_{j'} c(a_j, a_{j'}, \bm{x}, v) \\
&&- \sum_{m: a_m \in \mathscr{A} \setminus\{a_j, a_{j'}\} } p_m \lbc c(a_{j'}, a_m, \bm{x}, v) - c(a_j, a_m, \bm{x}, v) \rbc.
\end{eqnarray*}

We now show how we arrive at equation (6) for adjusting the responses so that the bias in equation (5) due to individual-level unmeasured confounding will be effectively removed from the confounding function adjusted treatment effect estimate.  

Because the causal effect is defined as the between-group difference in mean potential outcomes and is estimated based on the observed outcomes.  To correct the bias in equation (5) due to individual-level unmeasured confounding, we adjust the actual survival time $T$ of an individual who received treatment $a_j$ as 
\begin{eqnarray}\label{eq:tcf}
\log T^{CF} &=& \log T - \lsq E \lbc \log T(a_j) \cond a_j, \bm{x}, v\rbc -  E \lbc  \log T(a_j) \cond \bm{x}, v \rbc \rsq. 
\end{eqnarray} 

Let $p_j \equiv P(A=a_j \cond \bm{X}=\bm{x}, V=v)$ and $E \lsq \cdot \cond a_j, \bm{x}, v\rsq  \equiv E \lsq \cdot \cond A=a_j,\bm{X}=\bm{x}, V=v\rsq, \forall a_j \in \mathscr{A}$.  By applying the law of total expectation to $E \lsq  T(a_j) \cond \bm{x}, v \rsq$, we can rewrite the second quantity of the RHS of equation~\eqref{eq:tcf} as 
\begin{eqnarray*}
 &&E \lsq  \log T(a_j) \cond a_j, \bm{x}, v\rsq -  E \lsq  \log T(a_j) \cond \bm{x}, v\rsq \\
 &=& E \lsq  \log T(a_j) \cond a_j, \bm{x}, v \rsq  -  \sum_{m=1}^J p_m E \lsq  \log T(a_j) \cond a_m, \bm{x}, v\rsq \\
 &=&(1-p_j) E \lsq  \log T(a_j) \cond a_j, \bm{x}, v \rsq - \sum_{m\neq j}^J p_m E \lsq  \log T(a_j) \cond a_m, \bm{x}, v \rsq\\
 &=& \sum_{m \neq j}^J p_m \lbc E \lsq  \log T(a_j) \cond a_j, \bm{x}, v \rsq - E \lsq  \log T(a_j) \cond a_m, \bm{x}, v \rsq  \rbc\\
 &=&  \sum_{m \neq j}^J p_m c(a_j, a_m,\bm{x}, v).
\end{eqnarray*}
This implies that we will replace $T$ with $T^{CF}$ as $\log T^{CF} =\sum_{m \neq j}^J p_m c(a_j, a_m,\bm{x}, v)$,  which is equation (6). 

Now we prove that replacing $T$ with $T^{CF}$ removes the bias in equation (5). 
Consider the causal effect between any pair of treatments $a_j$ and $a_{j'}$. Using the adjusted survival times $T^{CF}$, the estimate of the causal effect is 
\begin{eqnarray*}
&&E \lsq  \log T^{CF} \cond a_j, \bm{x}, v  \rsq -E \lsq  \log T^{CF} \cond a_{j'}, \bm{x}, v  \rsq \\
&=& E \lsq \lp \log T - \sum_{m\neq j}^J p_m c(a_j, a_m,\bm{x}, v ) \rp \cond a_j, \bm{x}, v \rsq -  E \lsq \lp \log T - \sum_{m\neq j'}^J p_m c(a_{j'}, a_m, \bm{x}, v) \rp \cond a_{j'}, \bm{x}, v \rsq\\
&=& E (\log T \cond a_j, \bm{x}, v) - E (\log T \cond a_{j'}, \bm{x}, v)\\
&&\underbrace{+p_j c(a_{j'}, a_j, \bm{x}, v) - p_{j'} c(a_j, a_{j'}, \bm{x}, v) + \sum\limits_{m: m \in \mathscr{A}\setminus\{a_j, a_{j'}\}} p_{m} \lbc c(a_{j'}, a_m, \bm{x}, v) -c(a_j,a_m,\bm{x}, v) \rbc}_{\text{--Bias in equation (5)}}.
\end{eqnarray*} 

Because the survival time $T$ can be right censored, we can replace the centered complete-data survival time $\log y_{ik}^{cent,c}$ used in our riAFT-BART sampling algorithm for treatment effect estimation with adjusted $\log y_{ik}^{CF}$ as in equation (6). Running riAFT-BART with $\log y_{ik}^{CF}$ will effectively remove the bias in equation (5) from the confounding function adjusted treatment effect estimate. 

\section{Illustrative simulation for sensitivity analysis approach}

We considered a total sample size $N=10000 \; (n_k = 500, K=20) $ with an unbalanced treatment allocation (the ratio of units = 6:3:1) across three treatment groups, an individual-level binary measured confounder $X_{ik1} \sim \text{Bernoulli} (0.4)$, and an individual-level binary unmeasured confounder $X_{ik2} \sim \text{Bernoulli} (0.5)$. Both treatment assignment and outcome generating mechanisms depend on $X_{ik1}$ and $X_{ik2}$, but only $X_{ik1}$ is observed. The treatment assignment mechanism follows the following random-intercept multinomial logistic regression model, 
\begin{equation}\label{eq:trt_assign}
\begin{split}
\ln  \dfrac{P(A_{ik}=1)}{P(A_{ik}=3)} &= 1.6 + 0.2X_{ik1} + 0.4X_{ik2} + \tau_k \\
\ln  \dfrac{P(A_{ik}=2)}{P(A_{ik}=3)} &= -0.2 + 0.3X_{ik1} - 0.3X_{ik2} + \tau_k, \\
\end{split}
\end{equation}
where $\tau_k \sim N(0,1^2)$. Three sets of non-parallel response surfaces are generated as 
\begin{equation}
    T_{ik}(a_j)=
    \begin{cases}
      \lsq \dfrac{-\log U }{6\exp\lp-0.8X_{ik1}-1.2X_{ik2}+b_k\rp}\rsq^{1/2}, & \text{if}\ a_j = 1 \\
      \lsq \dfrac{-\log U}{2\exp\lp-0.5X_{ik1}-2.2X_{ik2}+b_k\rp} \rsq^{1/2}, & \text{if}\ a_j = 2 \\
      \lsq \dfrac{-\log U}{4\exp\lp -0.3X_{ik1}+1.0X_{ik2}+b_k\rp} \rsq^{1/2}, & \text{if}\ a_j = 3 \\
    \end{cases}
 \end{equation}
where $U$ is a random variable following the uniform distribution on the interval $[0,1]$ and $b_k \sim N(0,4^2)$. Observed and uncensored survival times are generated as $T_{ik} = \sum_{a_j \in \{1,2,3\}} T_{ik}(a_j)I(A_{ik}= a_j)$. Finally, we generate the censoring time $C$ independently from an exponential distribution with the rate parameter selected to induce 10\%  censoring. 

Under this simulation configuration, the true $\text{CATE}^0_{1,2}=0.31$, $\text{CATE}^0_{1,3}=0.21$ and $\text{CATE}^0_{2,3}=-0.10$ in terms of the log survival time in months.

\section{Supplementary Tables and Figures}

\begin{table}[H]
\centering
\caption{Interpretation of assumed priors on $c(a_j, a_{j'}, \cond \bm{x}, v)$ and $c(a_{j'},a_j \cond \bm{x},v)$ for average treatment effect based on log survival time. }
\label{tab:cfun-inter}
\begin{tabular}{ccp{0.72\textwidth}}
\toprule
\multicolumn{2}{c}{Prior assumption}  & Interpretation and implications of the assumptions\\
$c(a_j,a_{j'} \cond \bm{x},v)$ &$c(a_{j'},a_j\cond \bm{x},v)$ &\\\hline
$>0$ &$<0$ &  Individuals treated with $a_j$ will on average have longer potential survival time to both $a_j$ and $a_{j'}$  than individuals treated with $a_{j'}$; i.e. healthier individuals are treated with $a_j$. \\
$<0$ &$>0$ & Contrary to the above interpretation, unhealthier individuals are treated with $a_j$. \\
$<0$ &$<0$ & The potential survival time to $a_j (a_{j'})$  is shorter among those who choose it than among those who choose $a_{j'} (a_j)$.  Thus, the observed treatment allocation between these two approaches is undesirable relative to the alternative which reverses treatment assignment for everyone. \\
$>0$ &$>0$ & Contrary to the above interpretation, the observed treatment allocation between these two approaches is beneficial relative to the alternative which reverses treatment assignment for everyone.\\ 
\bottomrule
\end{tabular}
\end{table}

\begin{table}[H]
\centering
\caption{Specifications of treatment assignment model ($A$-model) in equation (8) and outcome generating model ($T$-model) in equation (10); and  coefficients $\xi^{L}_1, \xi^{L}_2$ and $\xi^{NL}_1, \xi^{NL}_2$ of the  $A$-model and $\beta^{L}_{a_j}$ and $\beta^{NL}_{a_j}$ of the $T$-model, $a_j \in \{1,2,3\}$. We set $\xi_{01} = 0.9$ and $\xi_{02} = -1.0$ to generate the 6:3:1 ratio of unit across three treatment groups.}
\begin{tabular}{clcccccccccc}
\toprule
&  \multicolumn{2}{c}{$A$-model} && \multicolumn{3}{c}{$T$-model} \\
\cmidrule{2-3} \cmidrule{5-7}
Variables & $a_j = 1$ & $a_j = 2$  && $a_j = 1$ &$a_j = 2$ & $a_j = 3$  \\
\midrule
$x_1$ & 0 & 0 && 1.2 & 1.0 & 0.9  \\
$x_2$ & 0.3 & 0.3 && 1.0 & 0.8 & 0.9  \\
$x_3$ & 0 & 0 && 1.2 & 1.0 & 0.9  \\
$x_4$ & 0.5 & 1.2 && 0 & 0 & 0  \\
$x_5$ & 0.4 & 1.1 && 0 & 0 & 0  \\
$x_6$ & 0.2 & 0.9 && 1.2 & 1.0 & 0.4  \\
$x_7$ & 0.3 & 0.5 && 0.6 & 0.8 & 0.5  \\
$x_8$ & 1.1 & 0.6 && 0.4 & 0.4 & 0.3  \\
$x_9$ & 0.6 & 0.7 && 0.5 & 0.6 & 0.2  \\
$x_{10}$ & 1.2 & 0.6 && 0.6 & 0.6 & 0.1  \\
$x_1^2$ & 0 & 0 && 0.9 & 0.7 & 0.85  \\
$x_1^3$ & 0 & 0 && 0.4 & 0.5 & 0.3  \\
$x_2^2$ & 0.4 & 0.75 && 0.3 & 0.8 & 0.7  \\
$x_2^2x_5$ & 0.4 & 0.8 && 0 & 0 & 0  \\
$x_2x_4$ & 0.5 & 0.9 && 0 & 0 & 0  \\
$x_2^2x_4$ & 0.8 & 0.7 && 0 & 0 & 0  \\
$x_2x_3$ & 0 & 0 && 0.3 & 0.4 & 0.8  \\
$x_2x_5$ & 0.7 & 0.7 && 0 & 0 & 0  \\
$x_2x_3^2$ & 0 & 0 && 0.4 & 0.7 & 0.8  \\
$x_2x_4x_5$ & 0.5 & 0.4 && 0 & 0 & 0  \\
$x_1x_2x_3$ & 0 & 0 && 0.3 & 0.5 & 0.7  \\
sin$(2\pi x_1x_3)$ & -0.8 & 0 && 0 & 0 & 0  \\
sin$(\pi x_1x_3)$ & 0 & -0.5 && 0 & 0 & 0  \\
sin$(\pi x_4x_5)$ & 0 & 0 && 0 & 0.6 & 0.5  \\
sin$(2\pi x_4x_5)$ & 0 & 0 && 0.6 & 0 & 0  \\
\bottomrule
\end{tabular}
\end{table}

\begin{table}[H]
\centering
\caption{The coverage probability for three treatment effect estimates $\widehat{CATE}_{1,2}$, $\widehat{CATE}_{1,3}$ and $\widehat{CATE}_{2,3}$ based on 5-year survival probability under four data configurations: (proportional hazards vs. nonproporitonal hazards) $\times$ (10\% censoring proportion vs. 40\% censoring proportion).} 
\begin{tabular}{clccccccc}
\toprule
& &  \multicolumn{3}{c}{Proportional hazards} && \multicolumn{3}{c}{Nonproportional hazards}  \\
\cmidrule{3-5}  \cmidrule{7-9}
 Censoring \%  &Methods & $\text{CATE}_{1,2}$ & $\text{CATE}_{1,3}$ & $\text{CATE}_{2,3}$ & & $\text{CATE}_{1,2}$ & $\text{CATE}_{1,3}$ & $\text{CATE}_{2,3}$ \\
\midrule
\multirow{4}{*}{10\%}  & IPW-riCox &16.0 	&20.4 	 & 18.4 & & 7.6	&	12.4 &10.4 \\
& DR-riAH & 77.2	&78.0	 &77.6 && 76.0	&	76.4 & 75.2\\
&ri-GAPH & 79.4	&80.0	 & 80.4&& 77.2	&   79.2	 &  78.0\\
&riAFT-BART & 	95.6 &	94.8 & 94.8   && 93.6   & 93.6 & 94.0\\
\midrule
\multirow{4}{*}{40\%}  &IPW-riCox &14.0 	& 18.8 	 & 16.8 	&&  6.0	&	10.8 &8.4 \\
& DR-riAH  &73.2	&75.6	 & 74.4  && 73.6	&73.2	 & 72.0 \\
& ri-GAPH   &76.8	&78.8	 &  77.6 &&74.4	&77.6	 & 75.2 \\
& riAFT-BART &94.8	 &94.4	 & 95.2	&& 92.8	 & 93.2	 & 93.6\\
\bottomrule
\end{tabular}
\label{tab:CP}
\end{table}

\begin{table}[H]
\centering
\footnotesize
%\scriptsize
\caption{Descriptions of pre-treatment variables and hospital locations (clusters) for each of three treatment groups in NCDB data.}
\begin{tabular}{lcccc}
  \hline
 & Overall & RP  & EBRT+AD  & EBRT+brachy$\pm$AD \\ 
 Characteristics & $N = 23058$ & $N = 14237$ & $N = 6683$ & $N = 2138$ \\
 \hline
 Age (years), mean (SD)& 71.14 (4.35) & 69.78 (3.23) & 73.69 (5.09) & 72.20 (4.42) \\ 
 Race, N ($\%$) &&&&\\
 \;\;White &19583 (84.9)   & 12335 (86.6)  &   5527 (82.7)   &   1721 (80.5)  \\ 
 \;\;Black & 3840 (14.1) &   1311 (9.2)    &    922 (13.8)  &    317 (14.8)   \\ 
 \;\;American Indian, Aleutian, or Eskimo &    925 (4.0)  &   591 (4.2)  &    234 (3.5)   &     100 (4.7)  \\ 
 Spanish or Hispanic Origin, N ($\%$) &&&&\\
  \;\; Non-Spanish; non-Hispanic & 22086 (95.8)  & 13649 (95.9)   &    6385 (95.5)   &   2052 (96.0) \\ 
 \;\; Spanish or Hispanic &972 (4.2) &  588 (4.1)  &  298 (4.1)  &   86 (4.0)  \\ 
 Insurance, N ($\%$) &&&&\\
 \;\;Yes & 22891 (99.3)& 14132 (99.3) & 6629 (99.2) & 2130 (98.8) \\ 
 \;\;No &    167 (0.7)  &   105 (0.7)  &     54 (0.8)   &  8 (0.4) \\ 
Income, N ($\%$) &&&&\\
 \;\;  $<$\$30,000 &  2464 (10.7) &  1462 (10.3)  &    755 (11.3)   &     247 (11.6) \\ 
 \;\;   \$30,000 - \$34,999 &  3936 (17.1)  &  2408 (16.9)  &   1171 (17.5)   &    357 (16.7)  \\ 
 \;\;   \$35,000 - \$45,999 &  6445 (28.0)  &  3977 (27.9) &   1927 (28.8)   &     541 (25.3)  \\ 
 \;\;   $>$\$46,000 &10213 (44.3)  &  6390 (44.9)  &  2830 (42.3)  &   993 (46.4) \\ 
 Education, N ($\%$) &&&&\\
\;\; $<$14\% & 9501 (41.2)  &  6161 (43.3)  &   2464 (36.9)  &    876 (41.0)   \\
\;\;    14\%-19.9\% &  5663 (24.6) &  3397 (23.9)  &  1753 (26.2)   &     513 (24.0)   \\
\;\;     20\%-28.9\% &  4838 (21.0) &  2852 (20.0)  &   1546 (23.1)  &    440 (20.6)  \\
\;\; $>$28.9\%     &  3056 (13.3) &  1827 (12.8) &    920 (13.8)  &    309 (14.5) \\
 Clinical T Stage, N ($\%$) &&&&\\
 \;\; $\leq$cT2 & 20509 (88.9)& 12972 (91.1) & 5699 (85.3) & 1838 (86.0) \\ 
 \;\; $\geq$cT3 &  2549 (11.1)  &  1265 (8.9)  &   984 (14.7)   &     300 (14.0)   \\ 
  Year of Diagnosis, N ($\%$) &&&&\\
 \;\; 2004-2010 &  2889 (12.5)  &  1825 (12.8)  &     670 (10.0)  &    394 (18.4)  \\ 
 \;\; 2011 &  3353 (14.5)  &   2069 (14.5)   &     859 (12.9)    &      425 (19.9)    \\ 
 \;\; 2012 &  3250 (14.1)  &   2048 (14.4)   &     892 (13.3)   &      310 (14.5)    \\ 
\;\; 2013 &  3949 (17.1)  &   2501 (17.6)   &   1165 (17.4)    &     283 (13.2)  \\ 
\;\; 2014 &  4292 (18.6)  &  2549 (17.9)   &    1418 (21.2)   &     325 (15.2)  \\ 
\;\; 2015 &  5325 (23.1)  &   3245 (22.8)   &   1679 (25.1)  &       401 (18.8)  \\ 
 PSA (ng/mL), mean (SD)& 17.11 (19.86) & 15.91 (19.01) & 19.77 (21.60) & 16.80 (18.90)\\ 
Gleason score, N ($\%$) &&&&\\
 \;\; 6 &    1016 (4.4) &   783 (5.5)  &       139 (2.1)    &        94 (4.4)   \\ 
 \;\; 7 &     3245 (14.1) &  1961 (13.8)   &      937 (14.0)    &        347 (16.2)   \\ 
 \;\; 8 &   11116 (48.2) &   7079 (49.7)  &     3041 (45.5)   &        996 (46.6)   \\ 
\;\; 9 &      7142 (31.0) &  4175 (29.3)  &   2312 (34.6)   &      655 (30.6)  \\ 
\;\; 10 &      539 (2.3) &   239 (1.7)   &      254 (3.8)   &          46 (2.2)    \\ 
Location, N ($\%$) &&&&\\
 \;\;    East North Central &  1020 (4.4)  &      755 (5.3)    &        177 (2.6)    &         88 (4.1)   \\ 
  \;\;    East South Central &  1850 (8.0) &     1367 (9.6)   &        312 (4.7)     &        171 (8.0)   \\ 
 \;\;    Middle Atlantic & 3558 (15.4)  &     1775 (12.5)    &      1469 (22.0)    &         314 (14.7)  \\ 
 \;\;    Mountain &  1594 (6.9)  &      731 (5.1)     &     822 (12.3)   &           41 (1.9)   \\ 
\;\;    New England &  4113 (17.8)  &     2521 (17.7)    &    1229 (18.4)     &        363 (17.0)   \\ 
\;\;    Pacific &  2910 (12.6)  &     2049 (14.4)     &        594 (8.9)    &          267 (12.5)    \\ 
 \;\;       South Atlantic &  4760 (20.6)  &     2520 (17.7)     &    1505 (22.5)    &      735 (34.4)  \\ 
 \;\;       West North Central &  2037 (8.8)  &       1566 (11.0)      &        353 (5.3)    &         118 (5.5)    \\ 
\;\;       West South Central &  1216 (5.3)  &     953 (6.7)  &       222 (3.3)    &         41 (1.9)   \\ 
   \hline
\end{tabular}
\footnotesize \\\smallskip
Abbreviations:  SD = standard deviation; PSA: Prostate-Specific Antigen
\end{table}

\begin{figure}[H]
\centering
\includegraphics[width = 1\textwidth]{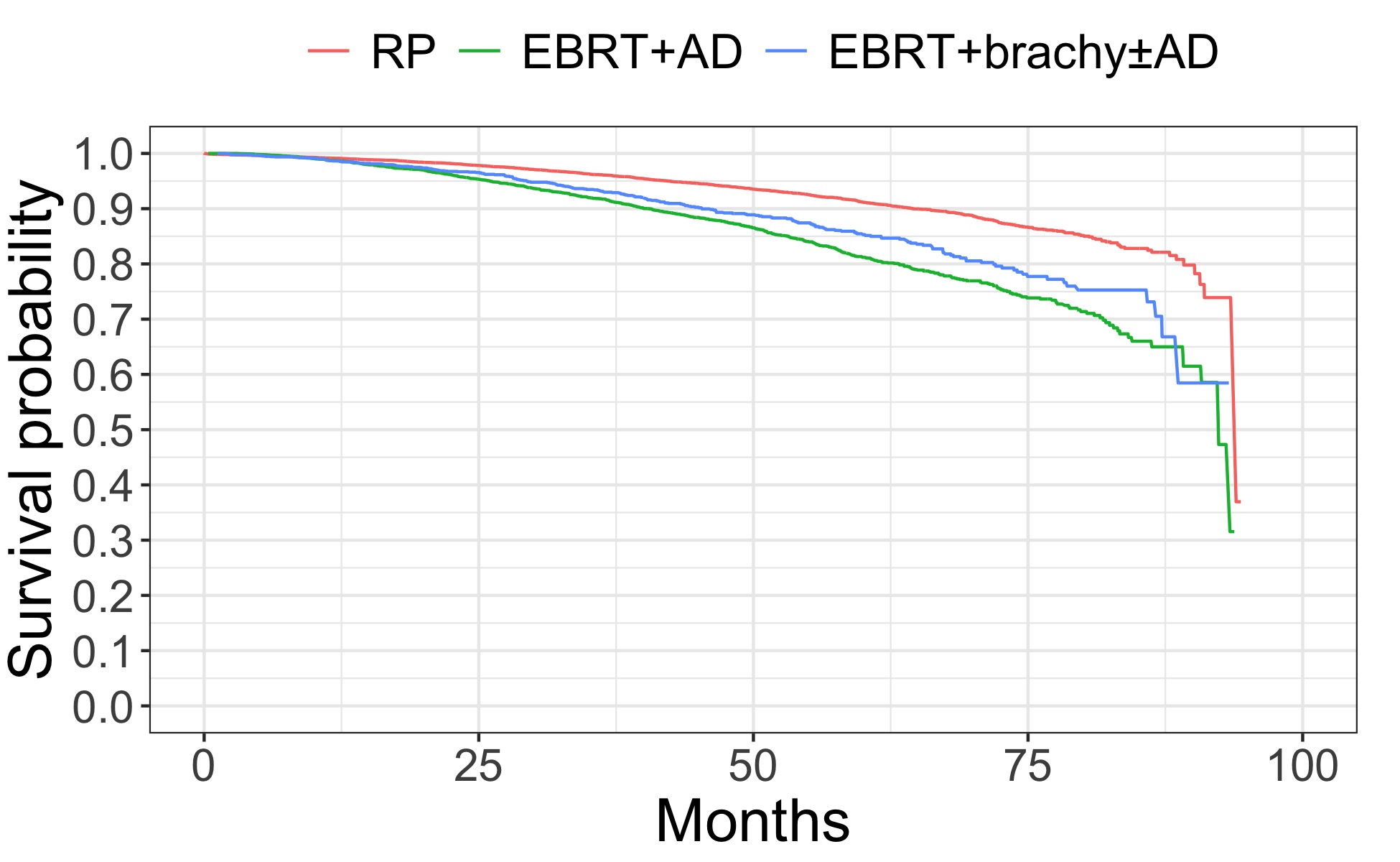}
\caption{The Kaplan-Meier survival curves for three treatment groups: radical prostatectomy (RP), external beam radiotherapy combined with androgen deprivation (EBRT+AD), and external beam radiotherapy plus brachytherapy with or without androgen deprivation (EBRT+brachy$\pm$AD), in NCDB data.}
\label{fig:KM-NCDB}
\end{figure}

\begin{figure}[H]
\centering
\includegraphics[width = 1\textwidth]{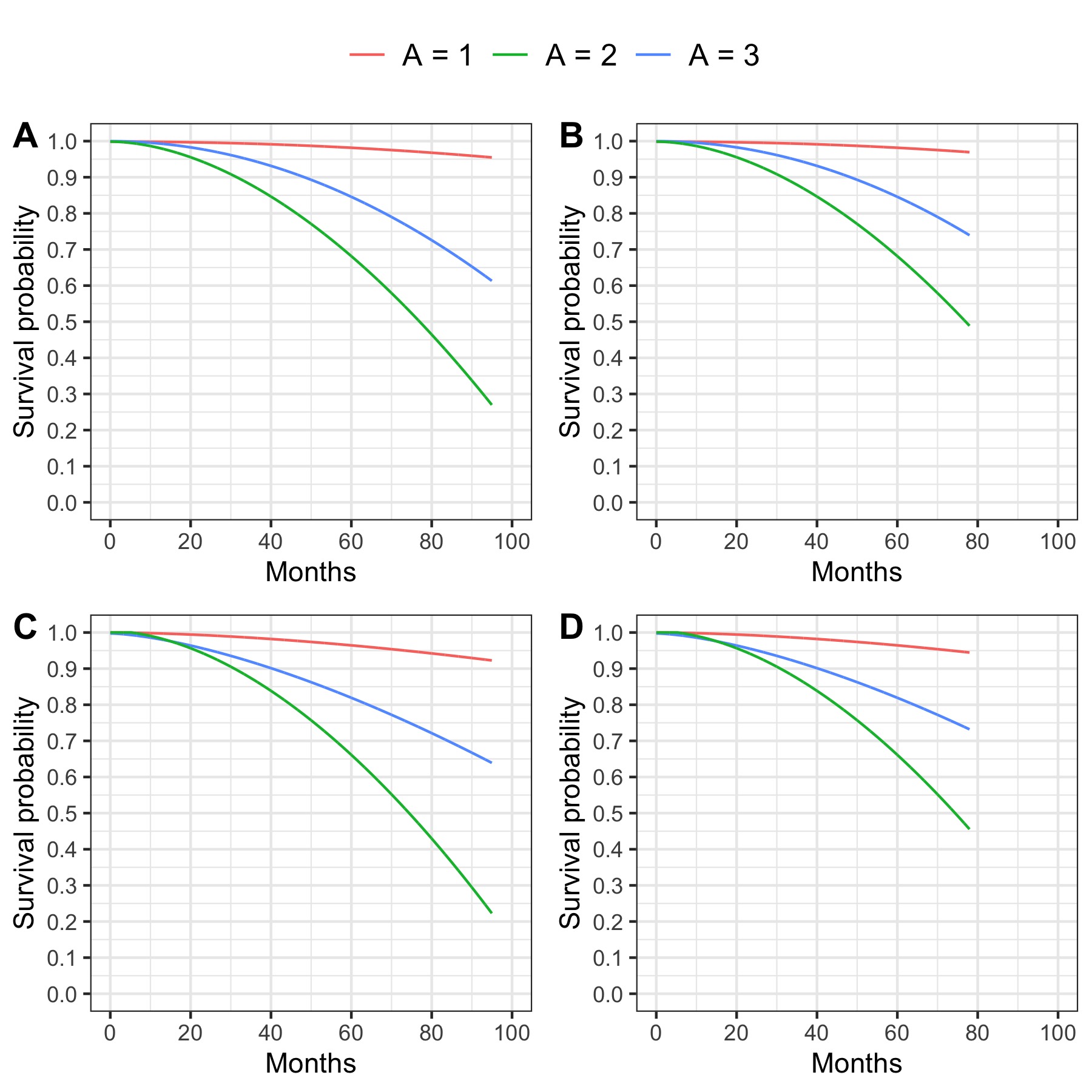}
\caption{Kaplan-Meier survival curves for three treatment groups generated in our simulation study in Section 4. Panels A--D respectively represent scenarios corresponding to proportional hazards with 10\% censoring, proportional hazards with 40\% censoring, non proportional hazards with 10\% censoring and  non proportional hazards with 40\% censoring.}
\label{fig:binary-sim}
\end{figure}

\begin{figure}[H]
\centering
\includegraphics[width = 1\textwidth]{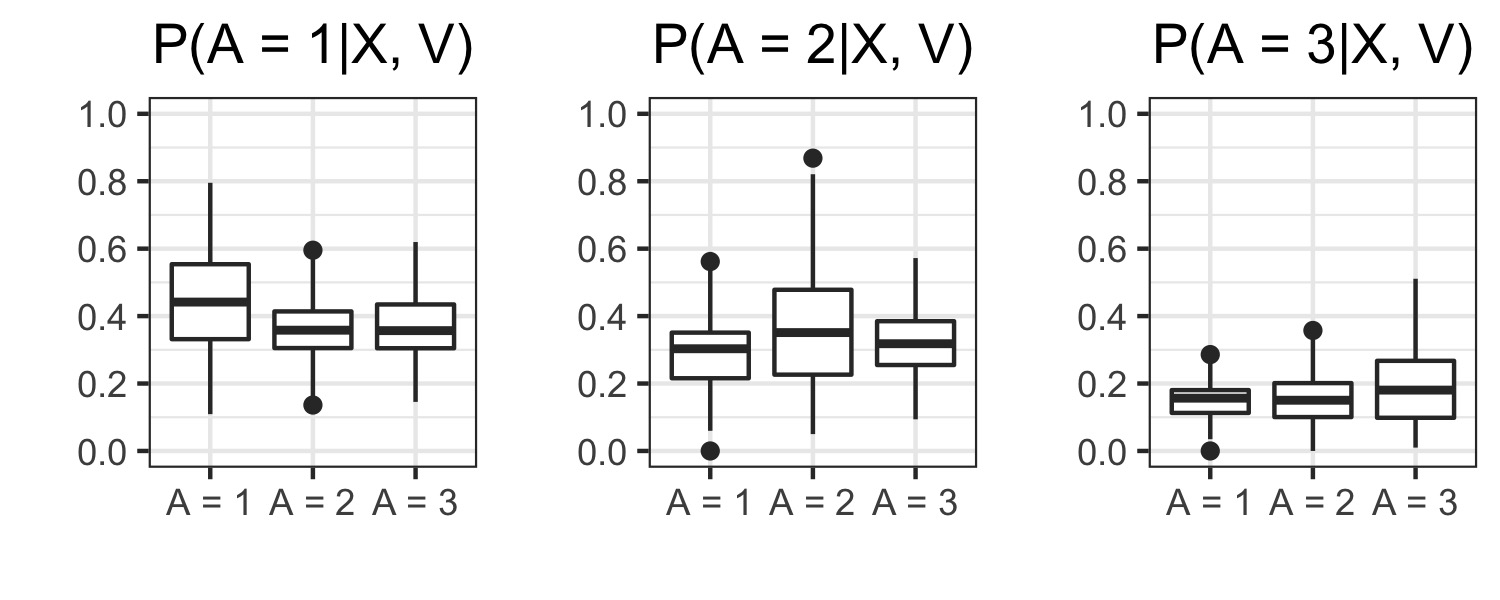}
\caption{Overlap assessment for data simulated under treatment assignment model in equation (8). Each panel presents boxplots by treatment group of the true generalized propensity scores for one of three treatments, and for every unit in the sample. The left panel presents treatment 1, the middle panel presents treatment 2, and the right panel presents treatment 3.}
\label{fig:overlap}
\end{figure}

\begin{figure}[H]
\centering
\includegraphics[width = 1\textwidth]{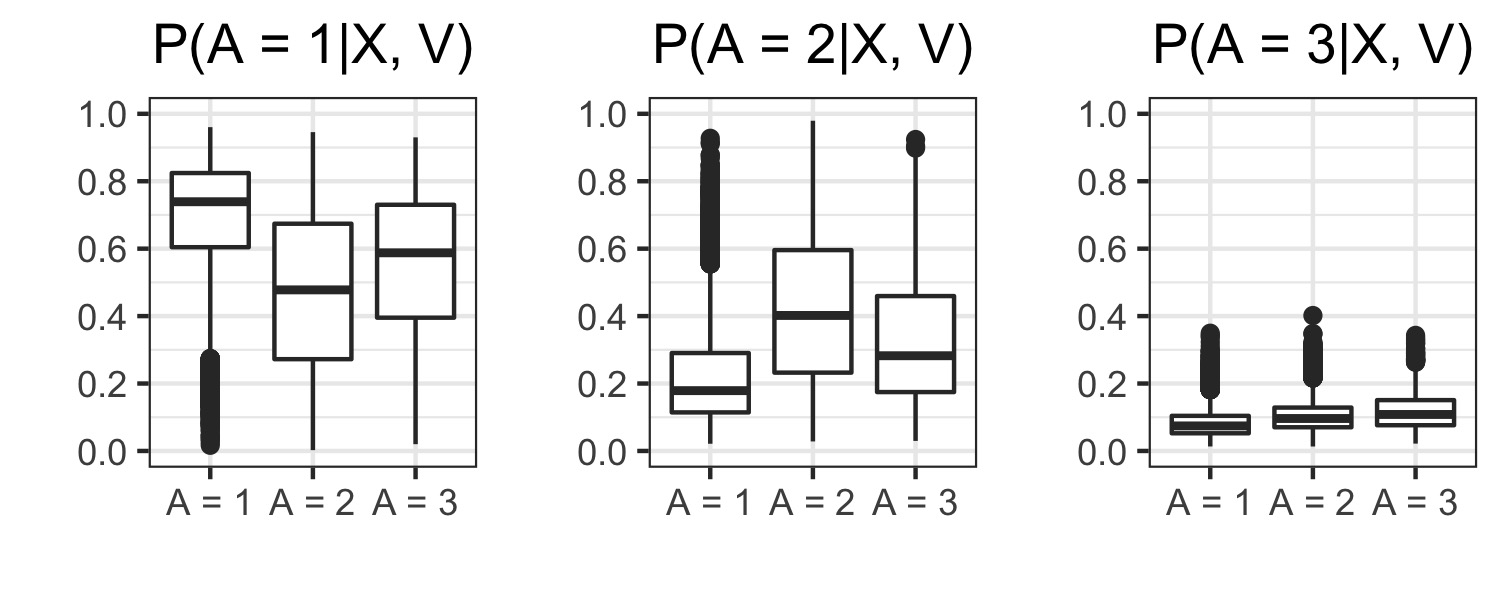}
\caption{Overlap assessment for three treatment groups in the NCDB data. Each panel presents boxplots by treatment group of the generalized propensity scores, estimated by Super Learner, for one of three treatments, and for every individual in the sample. The left panel presents treatment 1 = radical prostatectomy (RP), the middle panel presents treatment 2 = external beam radiotherapy combined with androgen deprivation (EBRT+AD), and the right panel presents treatment 3 = external beam radiotherapy plus brachytherapy with or without androgen deprivation (EBRT+brachy$\pm$AD).}
\label{fig:overlap}
\end{figure}

\begin{figure}[H]
\centering
\includegraphics[width = 1\textwidth]{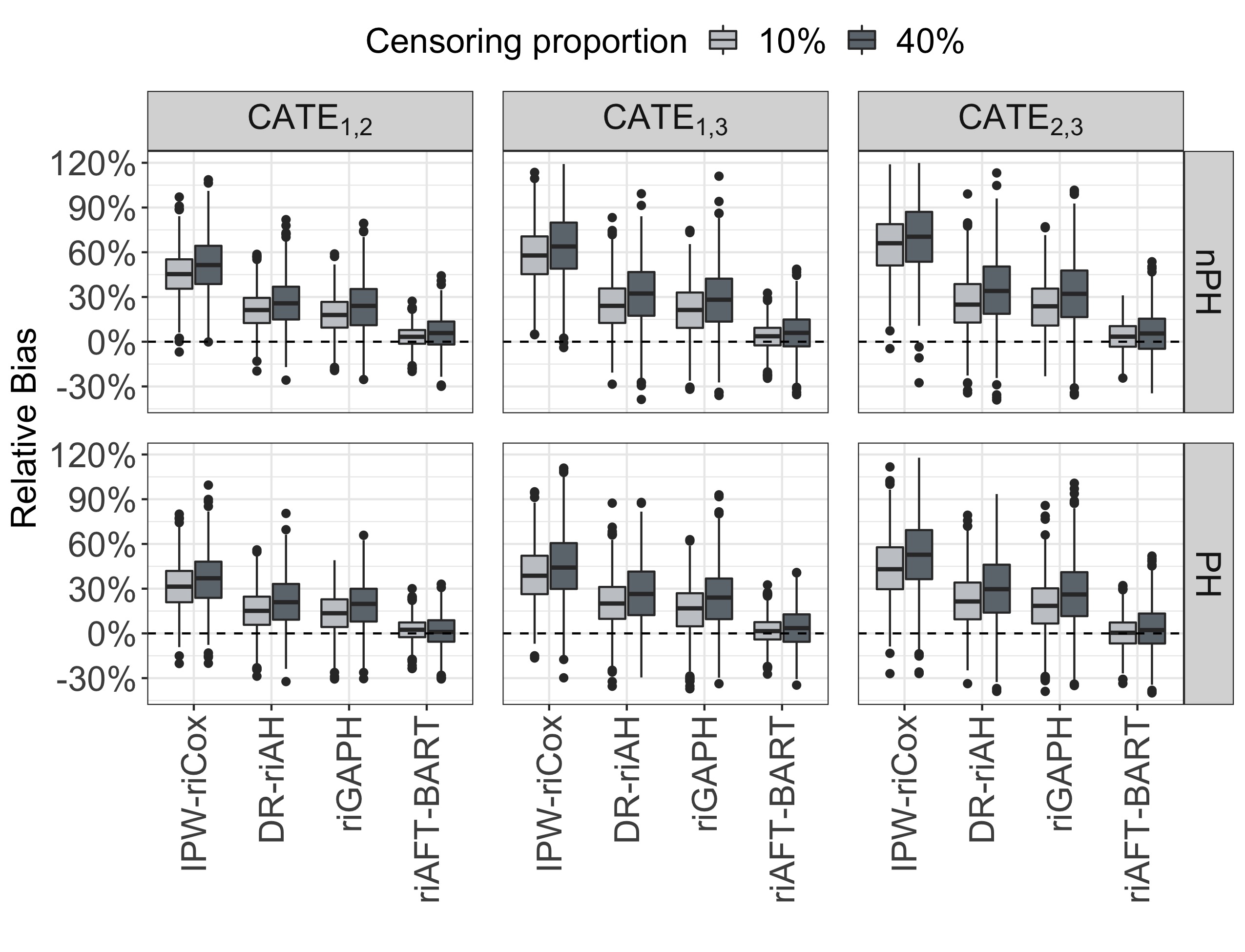}
\caption{Relative biases among 250 replications for each of four methods, IPW-riCox, DR-riAH, riGAPH and riAFT-BART, and three treatment effects  $CATE_{1,2}$, $CATE_{1,3}$ and $CATE_{2,3}$ based on 5-year survival under four data configurations: (proportional hazards vs. nonproporitonal hazards) $\times$ (10\% censoring proportion vs. 40\% censoring proportion). The true treatment effects under proportional hazards are $CATE^{0,PH}_{1,2} = 0.31$,  $CATE^{0,PH}_{1,3} = 0.16$ and $CATE^{0,PH}_{2,3} = -0.15$. The true treatment effects under nonproportional hazards are $CATE^{0,nPH}_{1,2} = 0.32$, $CATE^{0,nPH}_{1,3} = 0.17$  and $CATE^{0,nPH}_{2,3} = -0.15$.}
\label{fig:bias-surv}
\end{figure}

\begin{figure}[H]
\centering
\includegraphics[width = 1\textwidth]{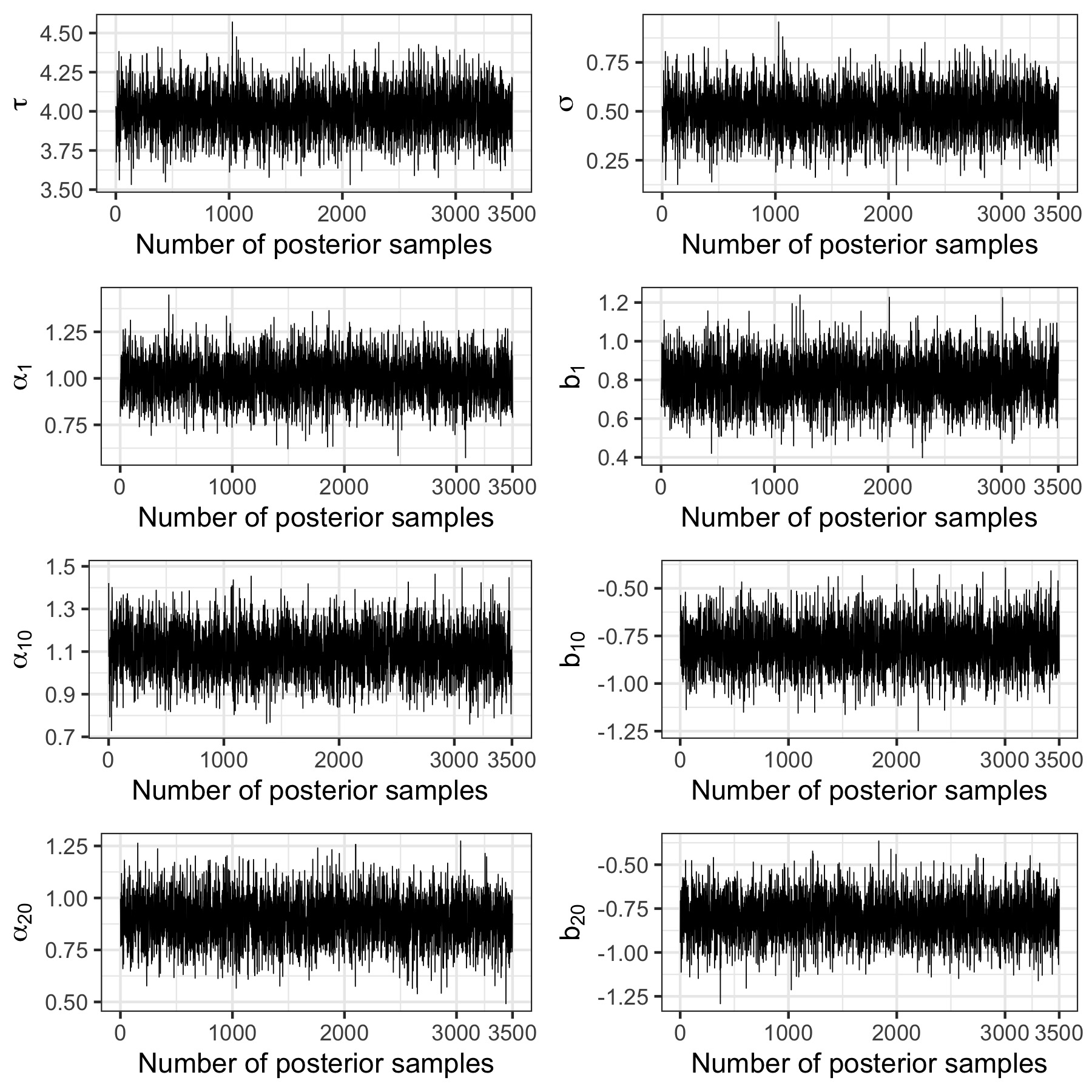}
\caption{Assessing convergence of the chain by plotting 3500 posterior draws of the variance parameters $\tau$ and $\sigma$, and cluster-specific parameter $\alpha_k$ and the random intercepts $b_k$ for clusters $k=1$, $k=10$ and $k=20$.}
\end{figure}

\bibliographystyle{biom}
\bibliography{reference}